\documentclass[11pt,a4paper]{article}
\usepackage[left=22mm,right=22mm,top=23mm,bottom=24mm]{geometry}
\usepackage{iftex}
\ifPDFTeX
  \usepackage[T1]{fontenc}
  \usepackage[utf8]{inputenc}
  \usepackage{lmodern}
\else
  \usepackage{fontspec}
\fi
\usepackage{amsmath,amssymb}
\usepackage{graphicx,booktabs,tabularx,array}
\usepackage[font=small,labelfont=bf,labelsep=period]{caption}
\usepackage{placeins}
\usepackage[numbers,square,sort&compress]{natbib}
\usepackage{xcolor}
\usepackage[unicode,breaklinks=true]{hyperref}
\hypersetup{colorlinks=true,linkcolor=black,citecolor=blue!45!black,
  urlcolor=blue!45!black,pdfauthor={Boliang Yu, Ruixin Zhou, Zi-Yu Tang},pdftitle={How tightly can PBH abundance information constrain scalar-induced gravitational waves?}}
\newcommand{\orcidlink}[1]{\href{https://orcid.org/#1}{\raisebox{-0.18ex}{\includegraphics[height=0.85em]{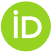}}}}
\let\savedthefootnote\thefootnote
\renewcommand{\thefootnote}{*}
\newcolumntype{L}{>{\raggedright\arraybackslash}X}
\newcolumntype{P}[1]{>{\raggedright\arraybackslash}p{#1}}
\title{How tightly can PBH abundance information constrain scalar-induced gravitational waves?}
\author{Boliang Yu\,\orcidlink{0009-0006-4664-5068}\textsuperscript{1}\thanks{\href{mailto:yuboliang@sjtu.edu.cn}{yuboliang@sjtu.edu.cn}}\quad Ruixin Zhou\,\orcidlink{0009-0004-9303-6674}\textsuperscript{1}\quad Zi-Yu Tang\,\orcidlink{0000-0003-3039-9975}\textsuperscript{2}\\[0.7em]
\parbox{0.96\textwidth}{\centering\small \textsuperscript{1} School of Physics and Astronomy, Shanghai Jiao Tong University, Shanghai 200240, China\\[0.25em]\textsuperscript{2} Cosmology, Gravity and Astroparticle Physics Group, Center for Theoretical Physics of the Universe, Institute for Basic Science, Daejeon 34126, Korea}}
\date{}
\begin{document}
\let\savedfootnotesize\footnotesize
\let\footnotesize\small
\maketitle
\let\footnotesize\savedfootnotesize
\let\thefootnote\savedthefootnote
\vspace{-2.5em}

\begin{abstract}

Primordial black hole (PBH) abundances and scalar-induced gravitational waves (SIGWs) respond differently to the same primordial curvature perturbations. For Gaussian perturbations, we find two single-peak lognormal spectra that satisfy the same synthetic abundance constraints over a finite stellar-mass interval, share an exact total abundance and obey external bounds. With a fixed quantum chromodynamics (QCD) formation prescription and ideal radiation for tensor evolution, their leading-order window-averaged SIGW energies differ by about 20\% in the lower of two adjacent nanohertz windows and by nearly a factor of 60 in the higher. Peak position, width and height compensate in the abundance fit, while almost all the higher-window energy comes from pairs of scalar modes with both wavenumbers above the primordial peak. Flexible spectra give larger contrasts at a fixed frequency under the QCD and stochastic peak-curvature prescriptions. Covering the mass-function turnover of a second synthetic population narrows the range of compatible lognormal shapes and reduces the higher-window contrast. Frequency-resolved SIGWs thus probe primordial shapes left unresolved by the PBH abundances.

\end{abstract}

\FloatBarrier

\section{Introduction}\label{sec:1}

Large primordial fluctuations can leave two different relics. Rare overdense regions may collapse into primordial black holes (PBHs), while couplings among scalar perturbations source a gravitational wave background in the early Universe \citep{young2024,domenech2024}. The latter signal is generated by the primordial fluctuations, independently of subsequent PBH mergers. Their common origin makes abundance measurements a possible route to predicting a companion scalar-induced gravitational wave (SIGW) spectrum \citep{saitoyokoyama2009}.

\begin{figure}[!htbp]
\centering
\includegraphics[width=\textwidth]{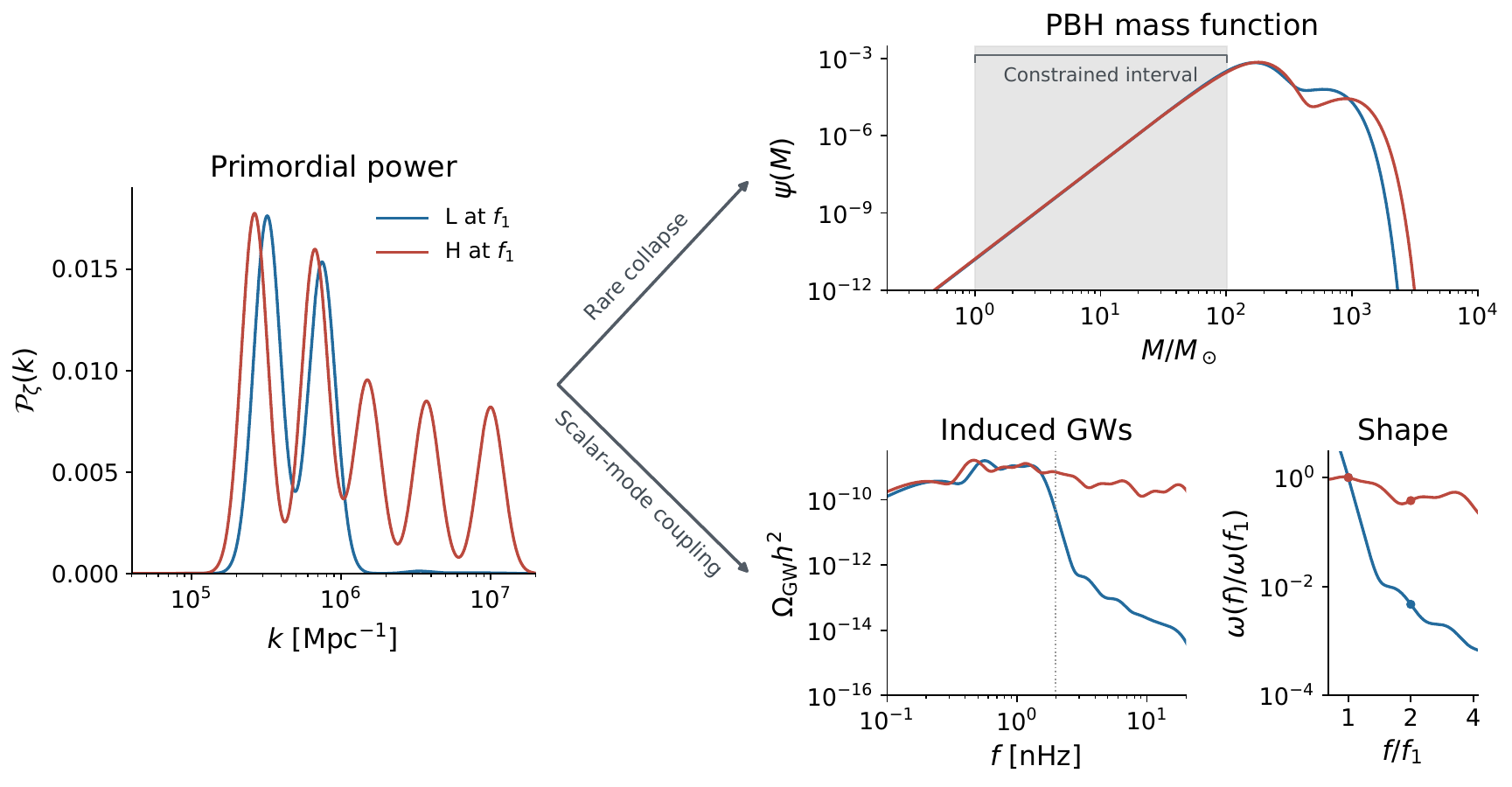}
\caption{Two predictions from the same primordial fluctuations, illustrated with the flexible QCD pair of Section 4.2. The left panel shows the curvature power spectra. The upper-right panel shows their PBH mass functions. They agree closely over the constrained mass interval (gray band and bracket) and separate on the high-mass side of the maximum. Both have the same total abundance. The lower-right panels show the leading-order tensor spectra and their shapes normalized at the reference frequency \(f_1=1.98\,\mathrm{nHz}\). Blue and red identify the lower (L) and higher (H) tensor signals at \(f_1\); dots in the shape panel mark the directly evaluated values at \(f_1\) and \(2f_1\). Both arrows start from the primordial perturbations.}
\label{fig:1}
\end{figure}

For Gaussian perturbations, the rarity of collapse makes PBH abundance exponentially sensitive to the smoothed fluctuation amplitude. At a specified primordial shape, precise abundance information can therefore fix the amplitude tightly. Smoothing and critical collapse broaden the map from curvature scale to PBH mass \citep{gowetal2021}, and thermal changes can imprint mass-function structure without corresponding primordial peaks \citep{escrivaetal2024}. Tensor production weights scalar pairs instead: resonance links two scalar scales to a tensor frequency, and the shapes and widths of scalar features affect the resulting spectrum \citep{kohriandterada2018,caietal2019,pisasaki2020,fumagallietal2021sharp}. Figure~\ref{fig:1} illustrates these two predictions using the flexible QCD pair analyzed in Section 4.2.

Finite peak widths affect the PBH--SIGW correspondence \citep{saitoyokoyama2010}. In the family studied by Bugaev and Klimai, a lower scalar peak height partly offsets the tensor growth from broadening under PBH constraints, leaving the maximum allowed amplitude weakly dependent on width \citep{bugaevklimai2011}. We examine how the frequency dependence above the tensor maximum can vary when mass-resolved abundance conditions are fixed. PTA searches and other SIGW analyses probe small-scale curvature power \citep{dandoyetal2023,franciolinietal2023,inomatanakama2019}. Following Kimura et al. \citep{kimuraetal2021}, Wang et al. \citep{wangetal2023reconstruction} reconstructed that power from selected GWTC-1 through GWTC-3 events under a PBH interpretation. Such reconstructions \citep{frolovskyetal2024,dingetal2026,elgammaletal2025} motivate quantifying what abundance information alone predicts.

We ask how much tensor freedom remains after fixing correlated abundance bins, an exact total and external population bounds. A smooth symmetric single peak provides the minimal setting in which peak position and width can compensate in formation. Allowing power to vary independently at separated scales then reveals how much larger the tensor alternatives can become and which scalar pairings produce them. Finally, extending the resolved mass distribution tests which additional abundance information constrains the remaining shapes.

The principal benchmark is a smooth synthetic population generated from a single localized curvature feature and normalized to prescribed abundance. Absolute information over a finite stellar-mass interval fixes the resolved mass distribution, an exact total fixes the population scale, and external weighted bounds supply information about the exterior distribution. These inputs define a conditional information model. Our principal constructions use Gaussian primordial perturbations and leading-order tensor spectra in ideal radiation.

An analytical amplitude elimination isolates the shape response that survives the exact total. Direct lognormal fits show unequal tensor responses in adjacent frequency windows. Flexible finite-width spectra extend this result through distinct mode couplings; changes in component width and a second formation prescription reveal how those couplings depend on the physical assumptions. A separate population generated directly by the QCD formation integral allows nested mass-coverage extensions, which distinguish the information in the lower mass interval from that near the turnover and body.

\FloatBarrier

\section{PBH formation and scalar-induced gravitational waves}\label{sec:2}

The dimensionless power spectrum \(\mathcal P_\zeta(k)\) describes the Gaussian curvature perturbation \(\zeta\) at comoving wavenumber \(k\). Each trial spectrum supplies two predictions, the induced tensor energy and the PBH mass function. We first give their common leading-order tensor calculation. For PBH formation, a collapse criterion selects overdense peaks and critical scaling assigns their masses. The minimal single-peak construction uses a fixed QCD formation calibration; the flexible class is evaluated with both that calibration and a rule integrating stochastic peak curvature. We denote the final PBH mass by \(M\) and its fraction of dark matter by \(f_{\rm PBH}\), with

\begin{equation}
\psi(M)\equiv\frac{df_{\rm PBH}}{d\ln M},\qquad
f_{\rm all}=\int_0^\infty\psi(M)\,d\ln M.
\label{eq:1}
\end{equation}

Here \(\psi\) is a fraction of the dark matter mass density per logarithmic mass. The mass weighting distinguishes it from the number distribution of individual PBHs.

\subsection{A common leading-order tensor calculation}\label{sec:2.1}

We compare the strain of freely propagating waves generated at second order in the scalar perturbations \citep{anandaetal2007,baumannetal2007}. Once the motion of the geodesic clocks is accounted for, the tensor contribution to their light travel time measures a gauge-invariant strain \citep{domenechpiwang2026}. At second order, a change of time slicing introduces terms quadratic in the scalar perturbations into the transverse-traceless metric perturbation \citep{domenechsasaki2018}. In Newtonian gauge the propagation strain is \(h_{ij}/2\) in the convention of \citep{kohriandterada2018}, where \(a^2h_{ij}/2\) is the transverse-traceless contribution to the spatial metric.

With the present scale factor \(a_0=1\), a comoving tensor wavenumber \(k\) corresponds to the observed frequency \(f=ck/(2\pi)\), where \(c\) is the speed of light. Well inside the horizon, we report the oscillation-averaged characteristic strain power \(h_c^2(f)\) as \(\Omega_{{\rm GW},0}(f)=2\pi^2 f^2 h_c^2(f)/(3H_0^2)\), the radiative energy density per logarithmic frequency in units of the critical density. Here \(H_0\) is the present Hubble rate and \(h=H_0/(100\,\mathrm{km\,s^{-1}\,Mpc^{-1}})\). For Gaussian seeds in ideal radiation, the spectrum is \citep{domenech2024,kohriandterada2018}

\begin{equation}
\omega(k)\equiv\Omega_{{\rm GW},0}(k)h^2
=\mathcal T_{\rm GW}
\int_0^\infty dv\int_{|1-v|}^{1+v}du\,
\mathcal K_{\rm RD}(u,v)
\mathcal P_\zeta(ku)\mathcal P_\zeta(kv).
\label{eq:2}
\end{equation}

The radiation kernel \(\mathcal K_{\rm RD}\) combines the Newtonian-gauge scalar source with the late-time oscillation average of the freely propagating tensor. Its arguments \(u,v\) are the two scalar wavenumbers in units of \(k\). The integration limits enforce the triangle condition. After the scalar source decays, the leading subhorizon spectrum agrees across gauges whose slicings satisfy the small-scale conditions of \citep{domenechsasaki2021}. At fixed frequency the energy and strain-power ratios coincide. Our logarithmic window averages weight the strain power by \(f^2\,d\ln f\).

We use \(\mathcal T_{\rm GW}=0.39\times4.18\times10^{-5}\), corresponding to the high-temperature degrees of freedom \(g_\rho=g_s=106.75\), as a common redshift convention. A common conversion cancels from H/L ratios. Appendix~\ref{app:B}.6 gives the strain normalization and timing response, and compares thermal evolution and scalar diffusion.

The convolution is quadratic in \(\mathcal P_\zeta\), so separated scalar features contribute through mutual coupling as well as self-coupling. The acoustic resonance near \(u+v=\sqrt3\) allows localized scalar features at \(k_a,k_b\) to enhance the tensor signal near \((k_a+k_b)/\sqrt3\) \citep{caietal2019}. Whether such power can be rearranged while preserving the PBH population requires the full nonlinear formation calculation. Appendix~\ref{app:B} gives the detailed calibrations and numerical evaluation.

\subsection{PBH formation with a fixed QCD calibration}\label{sec:2.2}

The first prescription uses the thermal degrees of freedom and QCD collapse tables accompanying Cecchini, Franciolini and Pieroni \citep{cecchinietal2025}, based on Franciolini et al. \citep{franciolinietal2022}. Here \(T\) is the radiation temperature, \(a\) the scale factor and \(H\) the Hubble rate. The effective entropy degrees of freedom are \(g_s(T)\); \(T_0\) and \(g_{s,0}\) are their present reference values. A comoving smoothing radius \(r_m\) enters the horizon when \(aHr_m/c=1\). With Newton's constant \(G\), the horizon mass \(M_H\) and horizon wavenumber \(k_{\rm hor}\) obey

\begin{equation}
\begin{aligned}
M_H(T)&=\frac{c^3}{2GH(T)},&
a(T)&=\frac{T_0}{T}\left[\frac{g_{s,0}}{g_s(T)}\right]^{1/3},\\
k_{\rm hor}(T)&=\frac{a(T)H(T)}{c},&
r_m&=k_{\rm hor}^{-1}.
\end{aligned}
\label{eq:3}
\end{equation}

The same temperature fixes the radiation density, horizon mass and collapse parameters. This connects a primordial length scale to the time and mass scale of formation.

The Gaussian compaction seed \(g\) measures the smoothed perturbation. The tabulated thermal factor \(\Phi(M_H)\) enters both its moments and the nonlinear compaction relation below. Its variance \(\sigma_c^2\) and second moment \(\sigma_{cc}^2\) are

\begin{equation}
\begin{aligned}
\sigma_c^2(M_H)&=\frac{4\Phi^2(M_H)}{9}
\int d\ln k\,\mathcal W(kr_m)\,\mathcal P_\zeta(k),\\
\sigma_{cc}^2(M_H)&=\frac{4\Phi^2(M_H)}{9}
\int d\ln k\,(kr_m)^2\mathcal W(kr_m)\,\mathcal P_\zeta(k),\\
\mathcal W(z)&=z^4W^2(z)W^2(z/\sqrt3),\qquad
W(z)=3\frac{\sin z-z\cos z}{z^3}.
\end{aligned}
\label{eq:4}
\end{equation}

The two top-hat factors describe smoothing and radiation transfer, respectively. Both moments are linear in \(\mathcal P_\zeta\). Their role in abundance is nonlinear because collapse selects the rare tail of the seed distribution. The effective count of high peaks per \(dg\) is \citep{iovinoetal2024}

\begin{equation}
p_{\rm pk}(g;\sigma_c^2,\sigma_{cc}^2)=\frac{b_{\rm pk}}{3\pi}
\left(\frac{\sigma_{cc}^2}{\sigma_c^2}\right)^{3/2}\frac{g^3}{\sigma_c^4}
\exp\!\left(-\frac{g^2}{2\sigma_c^2}\right),\qquad b_{\rm pk}=1.
\label{eq:5}
\end{equation}

This measure includes the peak-height Jacobian and is not a probability density normalized to unity. It does not integrate over the stochastic population of peak profiles \citep{pietal2025}.

The compaction \(\mathcal C\) measures the enclosed mass excess. Its collapse threshold is \(\mathcal C_{\rm th}\), while \(\kappa\) and \(\gamma\) are the normalization and exponent of the critical mass relation \citep{niemeyerjedamzik1998}. We retain the type I branch \(g<2\Phi\), on which compaction increases with the Gaussian seed,

\begin{equation}
\begin{gathered}
\mathcal C=g-\frac{g^2}{4\Phi},\qquad
M=\kappa M_H(\mathcal C-\mathcal C_{\rm th})^\gamma,\\
\mathcal C_{\rm th}<\mathcal C<\Phi,\quad g<2\Phi.
\end{gathered}
\label{eq:6}
\end{equation}

The four functions \(\kappa,\gamma,\mathcal C_{\rm th},\Phi\) are fixed QCD inputs when a candidate spectrum is varied in this prescription. Thus the thermal softening and mass dispersion are included, while the collapse calibration itself does not respond to spectral shape \citep{young2024,ianniccarietal2024}.

Let \(T_H\) and \(a_H=a(T_H)\) denote the temperature and scale factor associated with \(M_H\), \(\rho_E(T_H)\) the radiation energy density, and \(\rho_{{\rm DM},0}\) the present dark matter mass density. Weighting each peak by its final mass \(M_g\) and diluting its number density gives the present abundance; \(\delta_{\rm D}\) denotes the Dirac delta,

\begin{equation}
\begin{aligned}
\psi(M)&=\int d\ln M_H\,\frac{\rho_E(T_H)a_H^3}{\rho_{{\rm DM},0}c^2}
\int dg\,p_{\rm pk}(g;M_H)\frac{M_g}{M_H}
\delta_{\rm D}(\ln M-\ln M_g),\\
M_g&=\kappa M_H[\mathcal C(g)-\mathcal C_{\rm th}]^\gamma.
\end{aligned}
\label{eq:7}
\end{equation}

The inner integral is restricted to the collapsing type I domain in Eq. \eqref{eq:6}. We write \(\rho_{{\rm DM},0}=\Omega_{\rm DM}\rho_{c,0}\), where \(\Omega_{\rm DM}\) is the dark matter density parameter and \(\rho_{c,0}\) the present critical mass density. The formation energy fraction already includes \(M_g/M_H\). The mass weighting uses the same convention for the formation fraction as the monochromatic relation in \citep{carretal2021}.

\subsection{Amplitude rigidity and compensated shape changes}\label{sec:2.3}

We first use the QCD prescription to explain why rare-collapse sensitivity can make a fixed-shape tensor prediction very precise. Write \(\mathcal P_\zeta(k)=A S(k;\boldsymbol\theta)\), where \(A\) is a dimensionless amplitude and \(S\) specifies the shape through coordinates \(\boldsymbol\theta\), with its normalization convention fixed throughout each spectral family. For the lognormal family below, \(A=P_*\) is the peak height and \(S\) has unit peak height. Let \(Q_f(\boldsymbol\theta)\) be the tensor energy at frequency \(f\) evaluated with \(A=1\), and \(X\) any positive mass-weighted abundance integral in this prescription. At fixed shape, both moments in Eq. \eqref{eq:4} scale linearly with \(A\), whereas Eq. \eqref{eq:2} gives

\begin{equation}
\omega(f)=A^2Q_f(\boldsymbol\theta),\qquad
\frac{\partial\ln X}{\partial\ln A}
=\left\langle\frac{g^2}{2\sigma_c^2}-2\right\rangle_X
\equiv E_X .
\label{eq:8}
\end{equation}

The integral \(X\) may be a mass bin or the total, and the brackets denote its normalized formation weight. The large positive score of rare collapsing peaks makes \(E_X\) large, so a small amplitude change can strongly affect abundance. Appendix~\ref{app:A} gives the derivatives of the formation measure.

For a prescribed total \(f_{{\rm all},0}\), on a connected amplitude branch with \(E_{f_{\rm all}}>0\), the equation \(f_{\rm all}(A,\boldsymbol\theta)=f_{{\rm all},0}\) selects at most one amplitude for each fixed shape. If that spectrum passes the remaining conditions, its fixed-shape tensor prediction is unique on the branch. With a finite abundance error instead, the local propagation is \(d\ln\omega=2\,d\ln X/E_X\). This is the amplitude rigidity expected from rare-collapse sensitivity.

Shape changes need not inherit that rigidity. Solving the exact total for the amplitude at each shape gives the tensor response

\begin{equation}
\left.\frac{d\ln\omega(f)}{d\theta_j}\right|_{f_{\rm all}=f_{{\rm all},0}}
=\partial_{\theta_j}\ln Q_f
-2\frac{\partial_{\theta_j}\ln f_{\rm all}}
{\partial_{\ln A}\ln f_{\rm all}}.
\label{eq:9}
\end{equation}

The partial shape derivative holds \(A\) fixed. The first term reweights scalar pairs at unit amplitude; the second is the amplitude compensation required by the total abundance. Their sum is the physical tensor response. Changing the normalization convention for \(S\) changes this split while leaving the sum unchanged. Rare-collapse sensitivity does not force the terms to cancel, and different frequency windows can retain different balances between them. Appendix~\ref{app:B}.7 derives the abundance and window responses and relates the local derivatives to finite spectral changes.

\FloatBarrier

\section{Constructing spectra under common PBH constraints}\label{sec:3}

The principal benchmark specifies absolute abundance information in \(N_M=40\) logarithmic mass bins over \(\mathcal M_{\rm obs}=[M_-,M_+]=[1,101]\,M_\odot\), an exact all-mass total and external population bounds. The interval takes its endpoints from a mass reconstruction released with \citep{dingetal2026,dingwang2026data}. We choose the logarithmic bins and abundance errors for the synthetic measurement. The superscript 0 labels the reference population, and \(B_i^0\) denotes its abundance in bin \(i\).

The specified mass interval contains about 14\% of the reference population's total abundance and lies below the mass-function maximum. Its position relative to the turnover determines which shape information is available.

\subsection{Common absolute abundance conditions}\label{sec:3.1}

We ask how much tensor freedom remains when absolute PBH abundances are known to a specified precision over \(\mathcal M_{\rm obs}\). The observables are

\begin{equation}
\begin{gathered}
B_i[\mathcal P_\zeta]=\int_{M_i}^{M_{i+1}}\psi(M)\,d\ln M,
\qquad i=1,\ldots,N_M.
\end{gathered}
\label{eq:10}
\end{equation}

The target values \(\boldsymbol B^0\) obey \(f_{{\rm band},0}=\sum_iB_i^0\), and the total PBH fraction is fixed at \(f_{{\rm all},0}\). The principal lognormal and flexible-spectrum comparisons use these same inputs. The bins are not normalized within the observed band. Let \(\Sigma\) be the covariance of the log abundances. We choose a marginal log error \(\varepsilon=0.10\) for an informative synthetic measurement, with a correlation length motivated by critical-collapse mass dispersion. Appendix~\ref{app:B}.2 specifies the covariance model. Candidate populations must satisfy

\begin{equation}
\chi^2_{\rm PBH}[\mathcal P_\zeta]
=\bigl(\ln\boldsymbol B[\mathcal P_\zeta]-\ln\boldsymbol B^0\bigr)^{\mathsf T}
\Sigma^{-1}
\bigl(\ln\boldsymbol B[\mathcal P_\zeta]-\ln\boldsymbol B^0\bigr)\leq1.
\label{eq:11}
\end{equation}

The unit Mahalanobis ball is the declared common tolerance, rather than an \(N_M\)-dimensional 68\% confidence region. The exact total additionally prevents rescaling the overall population to explain the tensor difference; it leaves the exterior mass distribution unmeasured.

The target bins are generated from a smooth reference mass function based on a single lognormal curvature feature and Gaussian threshold statistics. We choose a subdominant population compatible with the adopted external bounds by normalizing it to the prescribed band abundance \(f_{{\rm band},0}=10^{-4}\); its full mass integral gives \(f_{{\rm all},0}\simeq7.25\times10^{-4}\). This normalization applies only to the target population. Every candidate uses its formation prescription's absolute mass-to-density conversion under the same abundance conditions. Appendix~\ref{app:B}.2 defines the generating spectrum, normalization and covariance. Section 4.5 uses a separate population generated directly by the absolute QCD integral to isolate the effect of extending mass coverage.

Three external PBH abundance bounds constrain the full population through cosmic radiation backgrounds, Galactic centre radio searches and cosmic microwave background (CMB) accretion feedback. Labeling them by \(a\), let \(f_{\max,a}(M)\) be the monochromatic abundance bound on tabulated mass support \(\mathcal I_a\). Its linear-response extension gives the population response \citep{carretal2017,bellomoetal2018},

\begin{equation}
\mathcal R_a[\psi]
=\int_{\mathcal I_a}d\ln M\,
\frac{\psi(M)}{f_{\max,a}(M)}\leq1.
\label{eq:12}
\end{equation}

We adopt the cosmic-radiation-background limits \citep{casanuevavillarrealetal2025}, Galactic-centre radio bound \citep{manshandenetal2019} and conservative Park--Ricotti CMB feedback bound \citep{agiusetal2024}. Equation \eqref{eq:12} assumes that the constrained signals add linearly over PBH number at fixed mass, using the published astrophysical conversions across the population \citep{carretal2017,bellomoetal2018}. We impose the three bounds separately on their tabulated supports. Appendix~\ref{app:B}.4 specifies the input curves; they are not direct measurements of \(\psi(M)\) or a joint accretion likelihood.

\subsection{A single lognormal peak}\label{sec:3.2}

A smooth symmetric peak allows its position and width to vary without introducing additional maxima or independent slopes on its two sides,

\begin{equation}
\mathcal P_\zeta(k)=P_*\exp\!\left[-\frac{\ln^2(k/k_*)}{2\Delta^2}\right].
\label{eq:13}
\end{equation}

Here \(P_*=\mathcal P_\zeta(k_*)\) is the peak height, \(\Delta\) is the standard deviation in \(\ln k\), and the integrated scalar power is \(\sqrt{2\pi}P_*\Delta\). Thus \(A=P_*\) in Section 2.3. Peak position and width vary, while the full QCD formation integral determines \(P_*\) from the exact total \(f_{\rm all}=f_{{\rm all},0}\). The target and absolute formation normalization remain fixed. Appendix~\ref{app:B}.3 gives the fitted parameters, and Appendix~\ref{app:B} summarizes the numerical accuracy.

\subsection{Tensor energy in adjacent frequency windows}\label{sec:3.3}

The NANOGrav 15-year analysis represents its low-frequency signal at Fourier harmonics \(i/T_{\rm NG}\), with data span \(T_{\rm NG}=16.03\,\mathrm{yr}\) \citep{nanograv2023}. We anchor our four comparison frequencies to its fundamental spacing,

\begin{equation}
\begin{gathered}
f_1=1.98\,\mathrm{nHz}\simeq T_{\rm NG}^{-1},\qquad
f_i=i f_1\quad(i=1,\ldots,4),\qquad
k_1=\frac{2\pi f_1}{c}\simeq1.28\times10^6\,\mathrm{Mpc}^{-1}.
\end{gathered}
\label{eq:14}
\end{equation}

We use the rounded Fourier spacing in Eq. \eqref{eq:14} as the numerical frequency input and derive each wavenumber through Section 2.1. The same spacing defines the point and window comparisons.

Within each pair, L and H denote the spectra with lower and higher tensor energy at \(f_1\). These labels remain fixed when the frequency changes.

We define equal-width intervals centered on the first two reference harmonics, with edges midway between adjacent harmonics,

\begin{equation}
\begin{aligned}
\mathcal B_j&=[(j-\tfrac12)f_1,(j+\tfrac12)f_1],\qquad j=1,2,\\
\overline\omega_{X,j}&=\frac{\int_{\mathcal B_j}\omega_X(f)\,d\ln f}
{\int_{\mathcal B_j}d\ln f},\qquad
\overline F_j=\frac{\overline\omega_{H,j}}{\overline\omega_{L,j}}.
\end{aligned}
\label{eq:15}
\end{equation}

We integrate each spectrum before taking the ratio. These windows diagnose finite-band energy differences around frequencies near the first two NANOGrav harmonics; they do not represent the sampling and timing-model response of a PTA data set.

The contrast of each constructed pair is \(F(f)=\omega_H(f)/\omega_L(f)\). At the reference frequency, \(F(f_1)\) is a constructive lower bound on the allowed dynamic range. Keeping the previously found spectra, further search can only lower the found minimum or raise the found maximum, so this bound cannot decrease. At other frequencies, L and H need not retain their ordering and \(F(f)\) can be less than one.

\FloatBarrier

\section{Results}\label{sec:4}

\subsection{Shape compensation in a single lognormal peak}\label{sec:4.1}

\begin{figure}[!htbp]
\centering
\includegraphics[width=\textwidth]{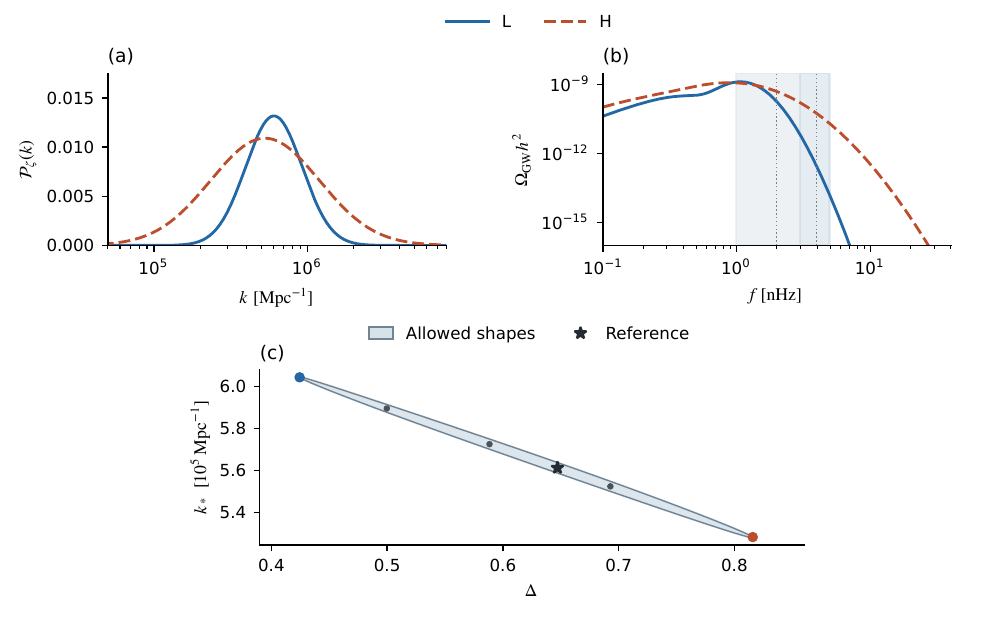}
\caption{Single lognormal spectra under the common abundance conditions. (a) Primordial power spectra. (b) Leading-order tensor spectra in ideal radiation, with the two adjacent integration windows shaded and dotted lines at \(f_1\) and \(2f_1\). (c) Peak-position and width compensation along the sampled abundance-compatible branch. The shading joins its sampled lower and upper boundaries; gray dots mark five branch samples and the star marks the reference fit. Colors identify L and H at \(f_1\). Bin-level abundance residuals are shown in Appendix~\ref{app:B}.3.}
\label{fig:2}
\end{figure}

Two smooth single lognormal spectra satisfying the common PBH conditions give window energy ratios \(\overline F_1\simeq1.2\) and \(\overline F_2\simeq58\). The corresponding higher-window means are \(\overline\omega_{L,2}\simeq1.2\times10^{-12}\) and \(\overline\omega_{H,2}\simeq7.2\times10^{-11}\), using the common ideal-radiation normalization of Section 2.1. The spectra have a contrast of about 2.7 at \(f_1\). H has a broader, lower peak at a slightly smaller wavenumber than L. Peak position, width and height compensate in the abundance calculation. Figure~\ref{fig:2} shows the spectra and their abundance-compatible shape branch. Appendix~\ref{app:B}.3 gives their parameters and constraint values.

The two frequency windows select different scalar pairs. Let \(q=ku\) and \(p=kv\) be the scalar wavenumbers sourcing a tensor mode \(k\). Their triangle condition is \(\lvert q-p\rvert\leq k\leq q+p\), so a tensor mode with \(k>2k_*\) requires at least one scalar wavenumber above the peak. The second window lies entirely above this monochromatic cutoff scale for both spectra; finite-width spectra retain power there \citep{pisasaki2020}. On the acoustic resonance \(q+p=\sqrt3k\), the triangle condition gives \(\min(q,p)\geq(\sqrt3-1)k/2\). Throughout the second window, both resonant scalar wavenumbers therefore lie above \(k_*\). Pairs with both wavenumbers above the peak supply more than 98\% of the second-window energy in each spectrum. In the first window, which samples the tensor peak region and straddles the two \(2k_*\) scales, their shares are about half and three quarters for L and H. The broader, shifted scalar peak changes the power sampled by the higher window much more strongly.

Formation weights the same power through the smoothed moments, the QCD collapse threshold and the critical mass relation. Partitioning the scalar contributions at twice each spectrum's peak makes their unequal weights quantitative. Pairs with both wavenumbers above \(2k_*\) supply about 60\% and 92\% of the second-window energies in L and H, whereas these modes contribute about 0.47\% and 6.0\% of the compaction variance averaged with the abundance weights. Each cut uses the spectrum's own peak, and the weights are its formation contributions to the observed PBH mass interval. This power participates in formation, but carries a much larger relative weight in the higher-window tensor energy. Changes in peak position, width and height preserve the mass-bin information and exact total through the full formation integral, while the higher-frequency tensor pairs retain a much larger change. Appendix~\ref{app:B}.3 defines the source fractions and abundance-weighted formation moments.

With peak height used as the amplitude, this finite compensation can be quantified directly. The unit-height tensor responses increase by factors of about 1.8 and 85 in the lower and higher windows. The common factor \((P_{*,H}/P_{*,L})^2\simeq0.69\) from the height change reduces them to the displayed energy ratios. Appendix~\ref{app:B}.7 relates the local formation and tensor responses to these finite spectral changes.

For this lognormal pair, scalar diffusion changes the second-window ratio from about 58 to about 53. The redistribution across frequency survives this change to the tensor source. Appendix~\ref{app:B}.6 gives the transfer comparison; absolute energies are provided with the supporting data.

\subsection{Flexible spectra under the QCD prescription}\label{sec:4.2}

To resolve how different scales share the formation and tensor responses, we also use a flexible finite-width spectral class. Write \(x=\ln[k/(1\,\mathrm{Mpc}^{-1})]\), with \(N_k\) components having centers \(x_\alpha\), widths \(\Delta_\alpha\) and nonnegative integrated areas \(c_\alpha\). Independent component areas allow local power rearrangements,

\begin{equation}
\mathcal P_\zeta(k)=\sum_{\alpha=0}^{N_k-1}
\frac{c_\alpha}{\sqrt{2\pi}\Delta_\alpha}
\exp\!\left[-\frac{(x-x_\alpha)^2}{2\Delta_\alpha^2}\right],
\qquad c_\alpha\geq0.
\label{eq:16}
\end{equation}

One component has the reference feature's center and width; the others share a logarithmic width \(\Delta_{\rm b}\). Under the same prescribed PBH population and QCD formation rule, the flexible pair gives \(F(f_1)\simeq15.5\), compared with 2.7 for the single lognormal. Appendix~\ref{app:B}.4 specifies the component parameters.

Over the common nanohertz interval \(\mathcal I=[0.1,100]\,\mathrm{nHz}\), define the integrated energies \(\mathcal E_X\) and total-variation distance \(D_{\rm TV}\),

\begin{equation}
\begin{aligned}
\mathcal E_X&=\int_{\mathcal I}\omega_X(f)\,d\ln f,\qquad X=L,H,\\
D_{\rm TV}&=\frac12\int_{\mathcal I}\left|\frac{\omega_H(f)}{\mathcal E_H}-\frac{\omega_L(f)}{\mathcal E_L}\right|\,d\ln f.
\end{aligned}
\label{eq:17}
\end{equation}

The ratio \(\mathcal E_H/\mathcal E_L\) compares integrated energy, while \(D_{\rm TV}\) measures shape after unit-integral normalization. The interval includes the tensor maxima and surrounding nanohertz structure. It defines a common spectral diagnostic rather than a detector sensitivity range; energy fractions and shape comparisons over wider and narrower intervals are supplied with the data.

Table~\ref{tab:1} compares these two QCD pairs on common frequency measures, together with the stochastic-curvature pair discussed in Section 4.4.

\begin{table}[!htbp]
\centering
\small
\setlength{\tabcolsep}{4pt}
\renewcommand{\arraystretch}{1.18}
\begin{tabularx}{\textwidth}{@{}P{0.29\textwidth}>{\raggedleft\arraybackslash}X>{\raggedleft\arraybackslash}X>{\raggedleft\arraybackslash}X@{}}
\toprule
Comparison & Lognormal, QCD & Flexible, QCD & Flexible, stochastic curvature \\
\midrule
First-frequency ratio \(F(f_1)\) & 2.7 & 15.5 & 11 \\
Integrated energy ratio \(\mathcal E_H/\mathcal E_L\) & 1.5 & 1.5 & 1.3 \\
Normalized shape distance \(D_{\rm TV}\) & 0.19 & 0.33 & 0.42 \\
H decline toward \(2f_1\) & Slower than L & Slower than L & Faster than L \\
First-window ratio \(\overline F_1\) & 1.2 & 1.7 & 1.1 \\
Second-window ratio \(\overline F_2\) & 58 & \(1.1\times10^3\) & 0.055 \\
\bottomrule
\end{tabularx}
\caption{The three constructed pairs under the common abundance conditions. The first two use the fixed QCD formation calibration; the third uses stochastic peak curvature. All tensor predictions here use ideal radiation at leading order. Integrated energies and normalized total variation use \(0.1\)--\(100\) nHz and measure \(d\ln f\). The last two rows average each energy spectrum separately in \(\mathcal B_1=[0.5,1.5]f_1\) and \(\mathcal B_2=[1.5,2.5]f_1\), then take H/L. The labels L and H always refer to \(f_1\).}
\label{tab:1}
\end{table}

This larger pointwise contrast also changes the frequency distribution. The integrated energy contrast over \(\mathcal I\) is much smaller than the value at \(f_1\), while the normalized spectra remain distinct (Table~\ref{tab:1} and Figure~\ref{fig:5}). H declines more slowly than L toward \(2f_1\), and the difference persists after averaging over finite windows. A single overall tensor normalization cannot represent these alternatives.

\begin{figure}[!htbp]
\centering
\includegraphics[width=\textwidth]{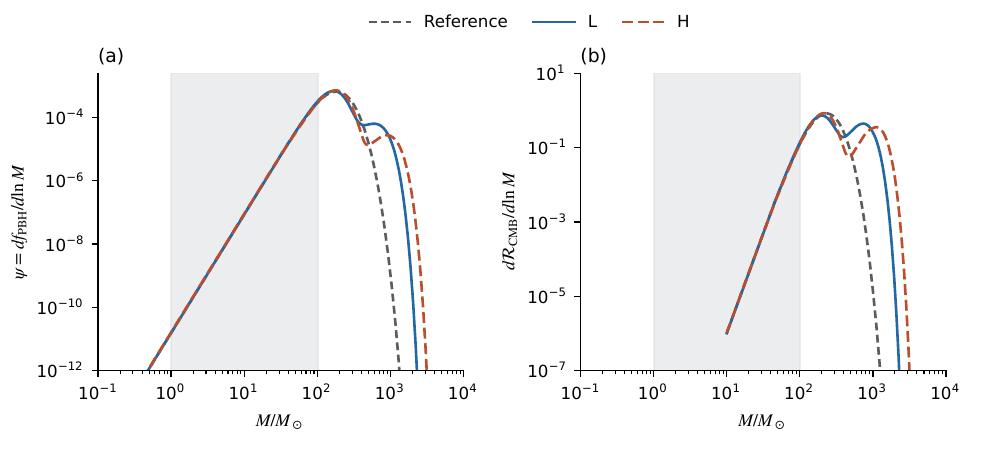}
\caption{Reference and flexible QCD populations. (a) Full PBH mass functions; shading marks the directly constrained mass interval. (b) CMB response density \(d\mathcal R_{\rm CMB}/d\ln M=\psi/f_{\max,\rm CMB}\), plotted only on the stated constraint-curve support. The CMB response emphasizes the population outside the observed interval. The abundance and external-bound values for L and H are given in Appendix~\ref{app:B}.4.}
\label{fig:3}
\end{figure}

The full mass functions show where the abundance information leaves room for compensation (Figure~\ref{fig:3}). About 86\% of the population lies outside \(\mathcal M_{\rm obs}\), predominantly between \(M_+\) and \(10^3M_\odot\). Together with the band abundances, the exact total constrains the exterior fraction but leaves its distribution in mass unresolved. The CMB response weights this exterior population strongly. Both flexible QCD populations have larger CMB responses than either minimal lognormal population. The larger realized spectral freedom therefore uses a different balance between the measured band and the CMB bound.

\begin{figure}[!htbp]
\centering
\includegraphics[width=\textwidth]{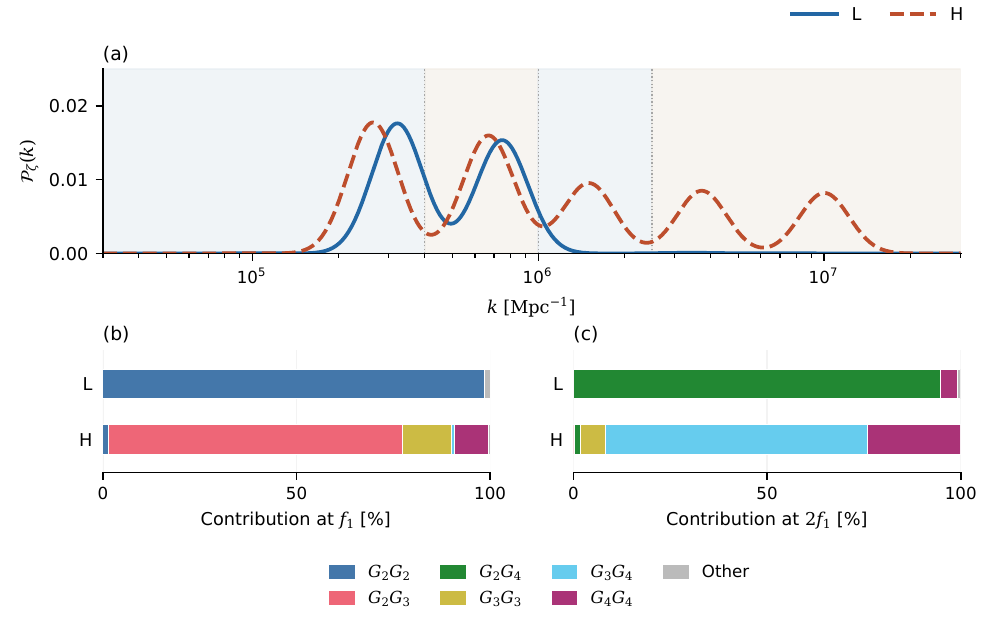}
\caption{Primordial power and tensor composition for the flexible QCD pair. (a) Curvature spectra; shading separates G1--G4 from left to right by component center, with component tails extending across the boundaries. (b,c) Fractional contributions at \(f_1\) and \(2f_1\). Each L or H bar is normalized to its own total signal at that frequency. Colors identify unordered scalar-group pairs: diagonal channels are \(D_{aa}\), and off-diagonal channels combine \(D_{ab}+D_{ba}\). Other sums all four unordered pairs involving G1. Absolute signal strengths are compared in Figure~\ref{fig:6}.}
\label{fig:4}
\end{figure}

For the mode decomposition, we divide component centers into four wavenumber groups G1--G4, shown by the shading in Figure~\ref{fig:4}. Each group retains its components' full tails. If \(\mathcal P_a\) is the scalar power in group \(a\), \(D_{ab}(f)\) denotes its ordered tensor-pair contribution with group \(b\), and \(\omega(f)=\sum_{a,b}D_{ab}(f)\). The abundance response to rescaling one group measures its formation weight. Appendix~\ref{app:A} defines the boundaries and derivatives.

The flexible QCD pair changes its dominant tensor pairing (Figure~\ref{fig:4}). At \(f_1\), G2 self-coupling supplies about 99\% of L, whereas G2--G3 coupling supplies about 76\% of H. For H, the G3 abundance elasticity is about 13 times smaller than that of G2 (Appendix~\ref{app:A}). At \(2f_1\), coupling between G3 and G4 instead dominates H, while L has already fallen to a weak tail. The same redistribution therefore changes both the amplitude and the frequency dependence. Removing G3 spoils the mass-bin fit, and restoring only the total abundance does not recover it. The H spectrum requires nonlinear compensation among its groups.

Scalar groups with unequal formation weights can participate in efficient tensor pairs. Coordinated changes in their power preserve the common PBH conditions while changing the tensor signal.

\subsection{Spectral width and resonant self-coupling}\label{sec:4.3}

Component width controls how much scalar power enters the efficient pairings identified above. We vary \(\Delta_{\rm b}\) at fixed centers and refit the component areas under the same QCD abundance conditions. Broadening mainly raises the low-signal member of the constructed QCD pairs. Increasing \(\Delta_{\rm b}\) from 0.20 through 0.30 to 0.40 gives constructed ratios 15.5, 5.7 and 3.5. The low signal rises by a factor of 3.8, whereas the high signal decreases by only 13\%. A comparably strong first-frequency signal remains realizable with broader components. The width sensitivity of the displayed contrast is concentrated in L.

The mode decomposition locates the low-signal increase in G2 self-coupling, which supplies at least 98\% of the energy at each width; G3 remains negligible. Most of the increase comes from scalar pairs near the acoustic resonance. Finite widths supply power above the G2 component centers, allowing their pairs to reach the acoustic resonance at \(f_1\). In the high spectra, the declining energy from G2--G3 coupling is partly replaced by increasing G3 self-coupling. Appendix~\ref{app:C}.2 defines the resonance neighborhoods and their contributions.

The unit-area Gaussians in Eq. \eqref{eq:16} give \(\int\mathcal P_\zeta(k)\,d\ln k=\sum_\alpha c_\alpha\). Holding the original low-end coefficients fixed and increasing \(\Delta_{\rm b}\) to 0.40 therefore preserves the total scalar area, yet raises the first-frequency tensor energy by a factor of 4.3. Refitting the PBH conditions reduces this to 3.8, offsetting about 11\% of the broadened signal. An overall amplitude adjustment alone fails to restore the abundance fit and CMB bound. Complete formation derivatives show that G2 remains strongly coupled to the band abundance. The power entering efficient tensor pairs therefore also participates in the compensation required by formation.

Reducing the number of local components retains a contrast of order ten, whereas broadening them primarily raises the constructed low signal (Appendix~\ref{app:C}.2).

\FloatBarrier

\subsection{Formation dependence}\label{sec:4.4}

The stochastic peak-curvature prescription of Germani and Sheth \citep{germanisheth2023} lets the candidate spectrum determine the distribution of peak shapes in a Gaussian field. It combines spatial peak counting with a scale-maximum condition and a radiation collapse threshold that depends on peak curvature. Critical scaling then assigns a mass to each collapsing peak. Appendix~\ref{app:B}.5 gives the complete statistics and mass integral.

\begin{figure}[!htbp]
\centering
\includegraphics[width=\textwidth]{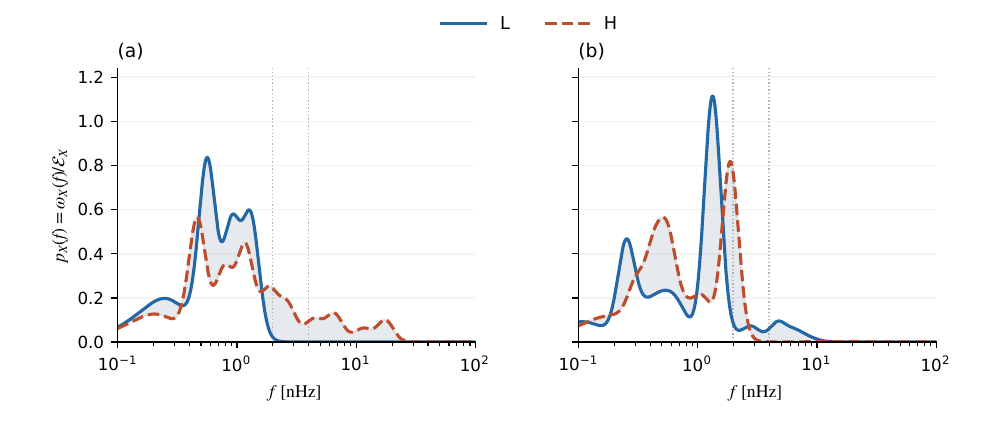}
\caption{Normalized leading-order tensor spectra in ideal radiation for (a) the flexible QCD pair and (b) the stochastic curvature pair. Each spectrum is divided by its own integrated energy \(\mathcal E_X\) over the displayed interval, so its integral in \(d\ln f\) is one. The shaded difference integrates to \(2D_{\rm TV}\), whose values are given in Table~\ref{tab:1}. Dotted lines mark \(f_1\) and \(2f_1\); colors identify L and H at \(f_1\).}
\label{fig:5}
\end{figure}

Under this rule, the same flexible class admits a different realization of tensor freedom. The target abundances, exact total and external bounds remain those of the principal comparison. The pair has a first-frequency contrast of about 11, with its common frequency measures listed in Table~\ref{tab:1}. Figure~\ref{fig:5} compares the normalized spectra under the two formation prescriptions.

The ordering reverses between the two flexible pairs. Under stochastic curvature, H falls faster than L and becomes the weaker spectrum at \(2f_1\), although it is stronger at \(f_1\). The continuous spectra cross between those frequencies. The QCD pair has the opposite decline ordering, as Table~\ref{tab:1} records.

\begin{figure}[!htbp]
\centering
\includegraphics[width=\textwidth]{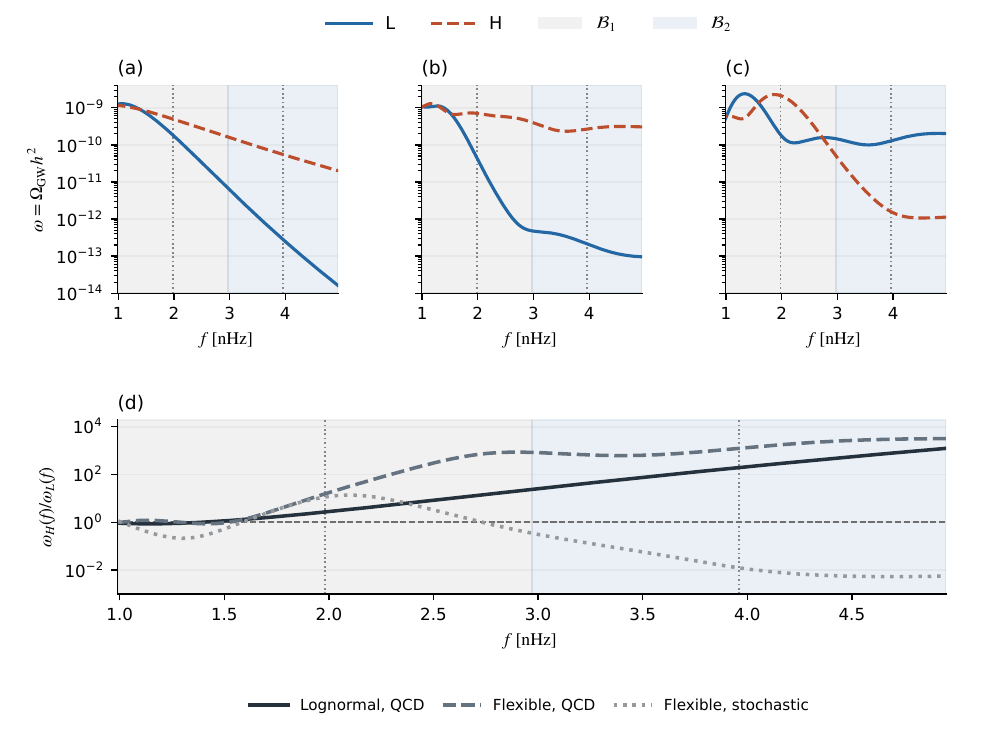}
\caption{Finite frequency windows for (a) the single-lognormal QCD pair, (b) the flexible QCD pair and (c) the stochastic curvature pair. Absolute ideal-radiation spectra use common axes, with \(\mathcal B_1\) and \(\mathcal B_2\) shaded. (d) Pointwise H/L ratios; the horizontal dashed line marks equal energy. Dotted lines mark \(f_1\) and \(2f_1\). The separately averaged window ratios are given in Table~\ref{tab:1}. L and H refer to the signal at \(f_1\); the stochastic curvature spectra exchange strength between the two frequencies.}
\label{fig:6}
\end{figure}

The localized H signal near 1.9 nHz makes its first-frequency contrast sensitive to bandwidth (Figure~\ref{fig:6}). Averaging each spectrum over the two stated windows preserves the exchange of strength, whereas integrating over the full interval gives a much smaller energy contrast (Table~\ref{tab:1}). These comparisons distinguish a large signal at one frequency from an overall amplification of the background.

\begin{figure}[!htbp]
\centering
\includegraphics[width=\textwidth]{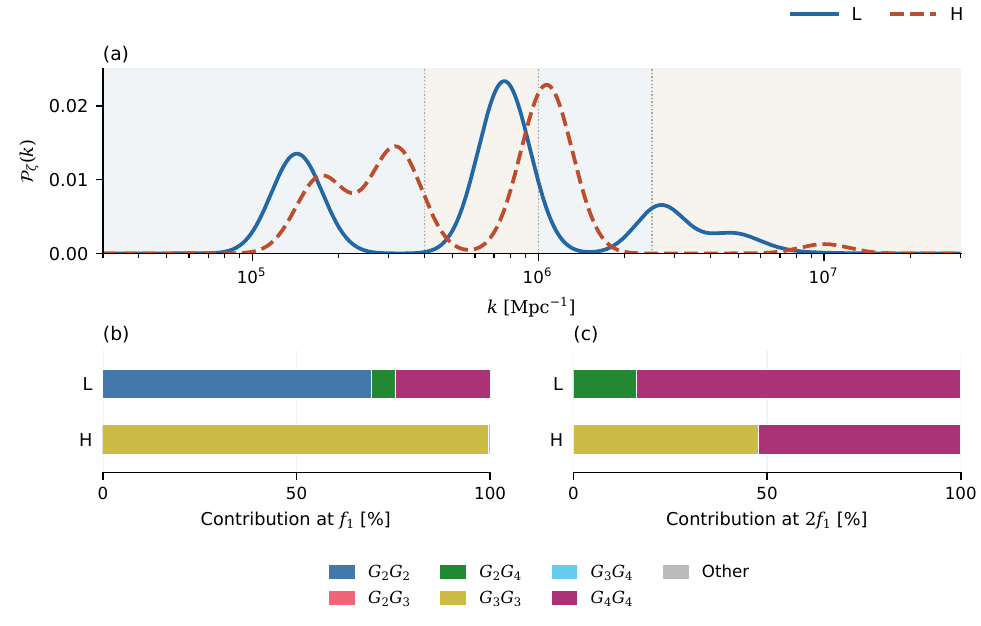}
\caption{Primordial power and tensor composition for the stochastic curvature pair. The panel order, group boundaries, axes and pair colors match Figure~\ref{fig:4}. Each L or H bar is normalized to its own total signal at the indicated frequency. At \(f_1\), G3 self-coupling dominates H. At \(2f_1\), G4 self-coupling supplies most of L, while G3 and G4 self-couplings contribute comparably to H. Other combines the four unordered pairs involving G1. Figure~\ref{fig:6} compares the absolute signals.}
\label{fig:7}
\end{figure}

The stochastic curvature pair realizes the contrast differently (Figure~\ref{fig:7}). G3 self-coupling supplies more than 99\% of its high signal at \(f_1\). This contribution falls sharply toward \(2f_1\), where the total is overtaken by L. Formation remains responsive to several groups: both G1 and G3 substantially affect H's abundance, even though only G3 dominates its first-frequency tensor signal (Appendix~\ref{app:A}). The negative G4 elasticity is the net change in the full peak statistics, not a negative population. The tensor-dominant component therefore does not independently determine the PBH abundance.

The window decomposition supports this interpretation. In the first window, G2 self-coupling dominates L in the stochastic pair and G3 self-coupling dominates H. In the second window, G4 dominates L while G3 still dominates H's much smaller energy. Formation and tensor production thus assign different weights to the same power in both prescriptions, but different pairs of scalar modes realize the allowed rearrangement.

With scalar diffusion, the stochastic pair's first-frequency ratio decreases from 11 to 8.4. Its H spectrum remains weaker than L in the second window, so the frequency ordering still reverses. Appendix~\ref{app:B}.6 summarizes this comparison; window averages and absolute energies are provided with the supporting data.

\FloatBarrier

\subsection{Mass coverage and shape compensation}\label{sec:4.5}

The remaining shape freedom also depends on which part of the PBH mass function is resolved. To isolate this information, we use a second target generated directly by the absolute QCD formation integral in Eq. \eqref{eq:7}. The same formation rule is applied to the reference lognormal and every candidate, with no rescaling of the resulting mass functions. Its peak position and width are those of the principal target's generating feature in Appendix~\ref{app:B}.2, and its height gives \(10^{-4}\) in the original mass interval. The resulting total \(f_{{\rm all},0}\simeq7.2\times10^{-4}\) and external bounds are fixed while mass information is added for this one population.

\begin{figure}[!htbp]
\centering
\includegraphics[width=\textwidth]{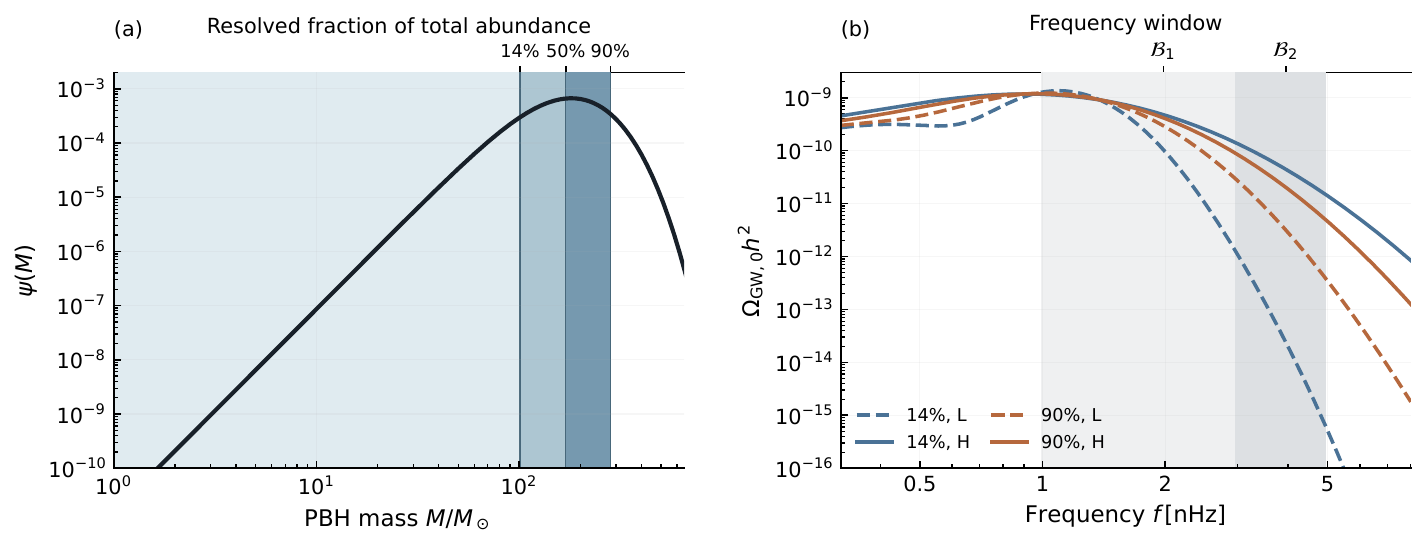}
\caption{Mass coverage and tensor predictions for the directly generated QCD population. (a) Reference mass function. The shaded regions mark successive extensions in final PBH mass. The upper axis gives the cumulative fraction of the reference abundance covered at each upper mass limit. (b) Leading-order tensor spectra for the pairs constructed with the original and 90\% coverages. Colors identify the coverage, and dashed and solid curves identify L and H at \(f_1\). Shading marks the two frequency windows. Table~\ref{tab:2} gives the window contrasts and the intermediate coverage.}
\label{fig:8}
\end{figure}

The original mass interval contains about 14\% of this population. Its abundance-weighted median mass is about \(170\,M_\odot\), and 90\% lies below about \(280\,M_\odot\). Extending the resolved abundances to these masses reaches the turnover and body of the mass function (Figure~\ref{fig:8}). The original bin information and external bounds are retained as bins are added; Appendix~\ref{app:C}.1 defines the extended covariance.

\begin{table}[!htbp]
\centering
\small
\setlength{\tabcolsep}{4pt}
\renewcommand{\arraystretch}{1.18}
\begin{tabularx}{\textwidth}{@{}L>{\raggedleft\arraybackslash}X>{\raggedleft\arraybackslash}X>{\raggedleft\arraybackslash}X@{}}
\toprule
Fraction of reference abundance covered & Upper mass \([M_\odot]\) & Lower-window excess \(\overline F_1-1\) & Higher-window ratio \(\overline F_2\) \\
\midrule
14\% & 101 & 25\% & \(3.2\times10^2\) \\
50\% & 170 & 13\% & 11 \\
90\% & 280 & 8.2\% & 4.2 \\
\bottomrule
\end{tabularx}
\caption{Window contrasts as mass information is added for the directly generated QCD population. Covered fractions integrate \(\psi(M)\,d\ln M/f_{{\rm all},0}\). Within each row, both window ratios use the pair selected by its low and high signals at \(f_1\); each spectrum is integrated before taking H/L. Each constructed ratio is a lower bound on the allowed contrast for that information model. The full spectrum parameters and nested abundance targets are available on Zenodo \citep{yuzhoutang2026dataset}.}
\label{tab:2}
\end{table}

With this common absolute formation rule, the original interval admits a higher-window contrast of about 320. Extending coverage through the median and then to 90\% of the abundance gives 11 and 4.2 for the constructed pairs (Table~\ref{tab:2}). At the generating spectrum, the smaller singular value of the covariance-weighted abundance Jacobian rises from 2.9 to 8.0, while the larger stays near 230.

Linearizing the bin abundances and window energies at the generating spectrum gives an independent estimate. At fixed total abundance, the unit abundance ellipsoid has half-width \(\Lambda_j\) in \(\ln\overline\omega_j\) (Appendix~\ref{app:B}.7), giving a ratio \(e^{2\Lambda_j}\) between its two extremes. In the higher window this gives 59, 9.9 and 4.3 in order of increasing coverage. The finite pairs give comparable contrasts at the two extended coverages. In the original interval the nonlinear branch reaches well beyond the linear estimate. In the lower window the linear estimates track the finite pairs at all three coverages.

The low-mass critical-collapse limit helps explain this information dependence. At a fixed formation scale, the leading tail has \(\psi(M)\propto M^{1+1/\gamma}\), with its shape set by the critical exponent. Finite-mass corrections carry more information about the primordial spectrum as the mass function approaches its turnover. These corrections already constrain both shape directions in the original interval. Extending through the body strengthens the weaker constraint and reduces the demonstrated tensor freedom, while leaving unequal responses in the two frequency windows.

\FloatBarrier

\section{Discussion}\label{sec:5}

\subsection{Implications for PBH--SIGW inference}\label{sec:5.1}

The single lognormal shows why an abundance fit need not select one tensor spectrum even without multiple primordial peaks. Moving and broadening the peak changes its mass-function response and the scalar pairs that source tensors; solving for the total abundance adjusts the amplitude but does not cancel the tensor change. The displayed flexible pairs realize larger contrasts through different mode couplings: between separated scalar groups in the QCD pair and mainly within one group in the stochastic-curvature pair. Their different frequency distributions show why a pointwise energy ratio cannot be treated as an overall enhancement of the background.

The frequency redistribution carries over to the radiative timing signal. The principal lognormal spectra give nearly equal timing RMS in the lower input window before timing-model fitting. Timing residuals weight low frequencies more heavily than the energy averages. In an ideal 16-year uniformly observed experiment with a quadratic timing trend removed, the broadband RMS ratio is about 1.17 (Appendix~\ref{app:B}.6). Matching the broadband RMS after the fit still leaves a factor of about six between the RMS contributions from the higher input window.

When the original mass bins and their errors are held fixed, adding bins through the turnover and body of the directly generated population constrains the combination of peak position and width that the low-mass tail leaves weakly determined. The higher-window contrast then falls much more than the lower-window contrast. The assumed abundance precision and the choice of target also affect their magnitudes (Appendix~\ref{app:C}.4).

Under the principal benchmark, additional high-mass abundance bins also reduce the constructed first-frequency contrast of the flexible spectra (Appendix~\ref{app:C}.1). Exterior mass information constrains power rearrangements that the original interval leaves weakly determined.

A narrower prediction needs information that distinguishes the remaining shapes, or an explicit restriction on the spectral family. Any inferred tensor lower bound should specify that family, the formation rule, exterior population bounds and the frequency observable. Reducing the number of spectral components still permits a substantial first-frequency contrast, while varying component width changes the constructed low signal. A separate broken-power-law test shows that single-peak spectra with independent slopes can also retain freedom (Appendix~\ref{app:C}.3).

Weak local non-Gaussianity can alter the amplitude and shape adjustments required by abundance matching, even when its explicit tensor terms are small. In the comparison of Appendix~\ref{app:B}.8, a two-window ratio alone does not identify whether the compensation comes from primordial statistics or spectral shape.

The CMB bound supplies another weight on the exterior population. Changing its mass dependence at fixed reference response retains substantial tensor contrasts. Its normalization also controls which populations remain compatible with the external limit, including the reference population itself (Appendix~\ref{app:C}.1). Frequency-resolved tensor information probes shape combinations that remain weakly constrained by the abundance bins. Broadband spectral shape can also distinguish tensor sources with matching peak frequencies and amplitudes, as shown by Maiti \citep{maiti2026spectralfingerprints} for scalar-induced and gauge-field-induced backgrounds.

\subsection{Scope of the physical models}\label{sec:5.2}

The two principal formation rules cover different approximations. The QCD prescription keeps its collapse calibration fixed \citep{franciolinietal2022,pietal2025,muscoetal2024}. The stochastic curvature prescription integrates candidate-dependent peak shapes using a local radiation threshold and a fixed critical mass coefficient \citep{germanisheth2023,escrivagermanisheth2020}; it does not evolve that population through QCD or include nonlocal peak overlap. Their differences in peak counting, thermal background and initial transfer convention prevent attributing numerical differences solely to profile dependence. The minimal lognormal calculation uses the QCD prescription; the independent stochastic-curvature construction uses the flexible class. Independent single-lognormal comparisons with joint threshold statistics retain unequal window responses (Appendix~\ref{app:B}.7). The local non-Gaussian comparison keeps the collapse inputs fixed; strong non-Gaussianity and a calibration of collapse for the resulting profile population lie beyond its scope.

The radiation prescription's CMB response includes formation scales in the keV era, where a fixed radiation fluid is not a kinetic treatment. For its displayed pair, \(M_H>10^6M_\odot\) supplies less than 0.2\% of either total but about 3.6\% and 24\% of the CMB responses. The CMB bound acts on final PBH mass, not on \(M_H\). Appendix~\ref{app:B}.5 separates these contributions from the negligible CMB weight of the formal extension beyond radiation domination.

All tensor observables reported here are leading-order predictions. Writing \(h_{[n]}\) for the tensor perturbation at order \(n\) in the primordial scalar field, the next order for Gaussian seeds contains both \(\langle h_{[3]}h_{[3]}\rangle\) and \(2\operatorname{Re}\langle h_{[2]}h_{[4]}\rangle\), including signed interference; the complete correction is not calculated. Ratios whose denominator lies in a weak spectral tail are particularly exposed to the uncomputed correction. The relative high-frequency third-order enhancement found for finite-width spectra in \citep{changetal2024} motivates this concern without determining a correction to these pairs. Assessing the higher-order accuracy of these predictions requires the complete tensor correction at that order.

For an isotropic, unpolarized tensor background, the radiative signal has the Hellings--Downs angular correlation between distinct pulsars \citep{nanograv2023}. Our calculation isolates this signal. Local scalar and vector perturbations, the motion of the clocks and the sampling and noise of a particular PTA lie outside the timing model.

The tabulated kernels of Franciolini, Pritchard and Tada include the Standard Model thermal history \citep{franciolinipritchardtada2026}. They show that the QCD crossover can enhance or suppress the SIGW spectrum in the PTA band, depending on when the dominant source modes enter the horizon. Our transfer calculation follows the crossover and joins ideal radiation below 5 MeV (Appendix~\ref{app:B}.6). The separate neutrino diffusion model uses a hydrodynamic damping ansatz extrapolated beyond parametrically small mean free paths \citep{yuetal2025,domenechandchluba2025}. Diffusion preserves the lognormal pair's large second-window difference and the stochastic pair's reversal in frequency ordering within that ansatz.

The synthetic covariance specifies a conditional information model rather than a measured catalogue error or formation uncertainty. The external curves are fixed response benchmarks rather than a joint accretion likelihood \citep{agiusetal2024}. Each feasible pair gives a lower bound on the allowed tensor contrast; the extrema of the full feasible set remain undetermined. The spectral families are phenomenological, with no inflationary dynamics specified.

\FloatBarrier

\section{Conclusion}\label{sec:6}

Precise PBH abundance information over a finite stellar-mass interval, an exact total and external bounds permit different leading-order tensor spectra even for one smooth lognormal peak. Peak position and width compensate with height in formation while reweighting the scalar pairs that generate tensors. For the principal pair, the lower window samples the tensor peak region, whereas the higher window draws almost all its energy from scalar modes above the primordial peak; their energy contrasts are about 20\% and a factor of 58. Flexible spectra realize larger pointwise differences through cross-coupling or self-coupling, with component width and formation statistics affecting the dominant pairings. Extending the mass information through the turnover and body of a directly generated QCD population restricts the peak-position and width compensation and reduces its higher-window contrast.

The predictive content of PBH information depends on which primordial shapes the resolved masses distinguish and how those shapes enter the tensor frequency response. Rare-collapse sensitivity tightly fixes amplitude at a specified shape, but precise abundances over a restricted interval need not determine the power sampled by higher-frequency scalar pairs. Measuring the mass-function body can restrict the remaining compensation; spectral assumptions and exterior bounds supply different additional information. Joint inference should propagate the primordial shapes compatible with the abundance information through the tensor spectrum and its frequency-dependent observational response.

\FloatBarrier

\section*{Acknowledgments}

We thank Qianhang Ding for helpful discussions. Z.-Y. Tang is supported by the Institute for Basic Science (IBS) under project code IBS-R018-D3.

\section*{Data availability}

The Zenodo dataset \citep{yuzhoutang2026dataset} contains the benchmark inputs, spectrum parameters, external bounds, PBH mass functions and tensor spectra. It also includes the nested mass-coverage targets, response decompositions, radiative timing data and local non-Gaussian comparisons. Appendix~\ref{app:B}.6 specifies the radiative timing model and the late-time QCD transfer calculation. The QCD formation inputs follow the fastPTA implementation \citep{cecchinietal2025}, and Section 3.1 specifies the external limits.

\FloatBarrier

\appendix

\section{Formation responses and group elasticities}\label{app:A}

For the flexible spectral class, we divide scalar components by their central wavenumbers into G1--G4: \(k<4\times10^5\), \([4\times10^5,10^6)\), \([10^6,2.5\times10^6)\), and \(k\geq2.5\times10^6\,\mathrm{Mpc}^{-1}\). These boundaries separate the populated component-center clusters of the displayed QCD pair and are then applied unchanged to the stochastic-curvature pair. Each group retains its components' full tails. Let \(\mathcal P_a\) be the power in group \(a\), and \(D_{ab}(f)\) the tensor contribution from ordered scalar pairs in groups \(a,b\). The total is \(\omega(f)=\sum_{a,b}D_{ab}(f)\), with both orders combined for contributions between different groups.

The formation response to the same group is its elasticity in the observed mass band,

\stepcounter{equation}
\begin{equation}
\begin{gathered}
E_a=\left.\frac{\partial\ln f_{\rm band}}{\partial\ln\lambda_a}\right|_{\lambda_a=1},
\qquad \mathcal P_a\mapsto\lambda_a\mathcal P_a,\\
f_{\rm band}=\sum_{i=1}^{N_M}B_i.
\end{gathered}
\tag{A1}\label{eq:A1}
\end{equation}

All other groups are held fixed in this derivative. The shape derivative in Eq. \eqref{eq:9} includes the amplitude change needed to preserve the total abundance. This elasticity measures the relative abundance response to changing one group. Coordinated changes in the full spectrum determine whether the common abundance conditions remain satisfied.

The ordered scalar-pair contribution is

\stepcounter{equation}
\begin{equation}
\begin{gathered}
D_{ab}(f)=\mathcal T_{\rm GW}\int dv\,du\,\mathcal K_{\rm RD}(u,v)
\mathcal P_a(ku)\mathcal P_b(kv),\\
\omega(f)=\sum_{a,b}D_{ab}(f).
\end{gathered}
\tag{A2}\label{eq:A2}
\end{equation}

Diagonal terms describe self-coupling; both orders are combined between different groups. Moving a group boundary by 10\% reassigns no component with appreciable area in the displayed QCD pair.

At fixed shape, \(\sigma_{cc}^2/\sigma_c^2\) in Eq. \eqref{eq:5} is independent of amplitude. The factor \(\sigma_c^{-4}\) contributes \(-2\) to its logarithmic derivative, and the exponential contributes \(g^2/(2\sigma_c^2)\). The collapse domain and mass weight remain fixed, giving Eq. \eqref{eq:8}.

The response of bin \(i\) to a logarithmic spectrum perturbation is defined by

\stepcounter{equation}
\begin{equation}
\delta\ln B_i=\int dx\,R_i(x)\,\delta\ln\mathcal P_\zeta(x)
+O\!\left((\delta\ln\mathcal P_\zeta)^2\right).
\tag{A3}\label{eq:A3}
\end{equation}

Here \(x=\ln[k/(1\,\mathrm{Mpc}^{-1})]\). These local derivatives are used for optimization and for the group elasticities; finite-amplitude alternatives are checked with the full nonlinear abundance calculation.

For the QCD formation prescription, set \(\mu_H=M_H/M_\odot\) and let \(\Pi_i\) be the normalized contribution density to mass-bin abundance \(B_i\) in the measure \(d\mu_H\,dg\), so that \(\int\Pi_i\,d\mu_H\,dg=1\). It includes the thermal weight, collapsed mass ratio and restriction to bin \(i\). The seed measure is the Gaussian density \(p_G\) in Eq. \eqref{eq:B3} for the threshold template and \(p_{\rm pk}\) for the peak prescription; the corresponding densities and responses carry superscripts \(G\) and \({\rm pk}\). Define the normalized spectral moment responses

\stepcounter{equation}
\begin{equation}
\begin{aligned}
r_c(M_H,x)&=\frac{[4\Phi^2/9]\,\mathcal W(kr_m)\,\mathcal P_\zeta(k)}{\sigma_c^2(M_H)},\\
r_{cc}(M_H,x)&=\frac{[4\Phi^2/9]\,(kr_m)^2\mathcal W(kr_m)\,\mathcal P_\zeta(k)}{\sigma_{cc}^2(M_H)}.
\end{aligned}
\tag{A4}\label{eq:A4}
\end{equation}

Each integrates to one in \(dx=d\ln k\). Direct differentiation of Eqs. \eqref{eq:B3} and \eqref{eq:5} gives

\stepcounter{equation}
\begin{equation}
\begin{aligned}
R_i^{G}(x)&=\int d\mu_H\,dg\,\Pi_i^{G}
\frac{g^2/\sigma_c^2-1}{2}\,r_c(M_H,x),\\
R_i^{\rm pk}(x)&=\int d\mu_H\,dg\,\Pi_i^{\rm pk}
\left[\frac{g^2/\sigma_c^2-7}{2}\,r_c(M_H,x)
+\frac32r_{cc}(M_H,x)\right].
\end{aligned}
\tag{A5}\label{eq:A5}
\end{equation}

Let \(g_{\rm th}\) be the Gaussian seed at the collapse threshold. The factor from the rare Gaussian tail is therefore not simply \(g_{\rm th}^2/(2\sigma_c^2)\); it is integrated over the masses, compactions and horizon scales contributing to the actual observable. The high-peak map has a different variance derivative and an additional spectral moment term. Equation \eqref{eq:A5} differentiates the formation model with the profile calibration and collapse threshold held fixed.

For the group scaling in Eq. \eqref{eq:A1}, \(\partial\ln\mathcal P_\zeta/\partial\ln\lambda_a=\mathcal P_a/\mathcal P_\zeta\). Weighting the bin responses by \(B_i/f_{\rm band}\) therefore gives

\stepcounter{equation}
\begin{equation}
E_a=\sum_{i=1}^{N_M}\frac{B_i}{f_{\rm band}}\int dx\,R_i(x)\frac{\mathcal P_a(k)}{\mathcal P_\zeta(k)}.
\tag{A6}\label{eq:A6}
\end{equation}

For the QCD prescription this derivative keeps the formation tables fixed and includes both moments. Its locality does not imply that a finite change in one group alone preserves the abundance conditions.

For stochastic curvature, the seed and scale derivatives are \(g=-(4/3)r\partial_r\zeta\), \(v=r\partial_r g\) and \(w_{\rm pk}=-r^2\partial_r^2g\), where \(r\) is the comoving smoothing radius. The spectral moments \(\sigma_1,\sigma_2\), characteristic radius \(R_*\) and Bardeen--Bond--Kaiser--Szalay (BBKS) peak function are defined in Appendix~\ref{app:B}.5. Let \(\mathbf C_{\rm pk}\) be the covariance of \((g,v,w_{\rm pk})\) and evaluate its Gaussian density at \(\boldsymbol y=(g,0,w_{\rm pk})^{\mathsf T}\). Define \(\xi=(2g+w_{\rm pk})/\sigma_2\). A variation of the curvature spectrum at fixed \((r,g,w_{\rm pk})\) gives the logarithmic change in the measure of Eq. \eqref{eq:B10},

\stepcounter{equation}
\begin{equation}
\begin{aligned}
\delta\ln n_{\rm pk}={}&\frac12\left[\boldsymbol y^{\mathsf T}\mathbf C_{\rm pk}^{-1}(\delta\mathbf C_{\rm pk})\mathbf C_{\rm pk}^{-1}\boldsymbol y-\operatorname{tr}(\mathbf C_{\rm pk}^{-1}\delta\mathbf C_{\rm pk})\right]\\
&-\frac32\delta\ln\sigma_1^2
+\left[\frac32-\frac{\xi f'_{\rm BBKS}(\xi)}{2f_{\rm BBKS}(\xi)}\right]\delta\ln\sigma_2^2.
\end{aligned}
\tag{A7}\label{eq:A7}
\end{equation}

Here \(n_{\rm pk}\) denotes the differential measure in Eq. \eqref{eq:B10}. The covariance and both squared moments are linear in the component powers. Integrating this score with the appropriate mass weight gives each bin, total and external response. At fixed \(w_{\rm pk}\), the common threshold and mass rule have no further explicit spectrum derivative. Equation \eqref{eq:A7} includes the covariance, \(R_*\) and BBKS contributions. The analytic derivatives agree with symmetric finite differences under step refinement.

\begin{figure}[!htbp]
\centering
\includegraphics[width=\textwidth]{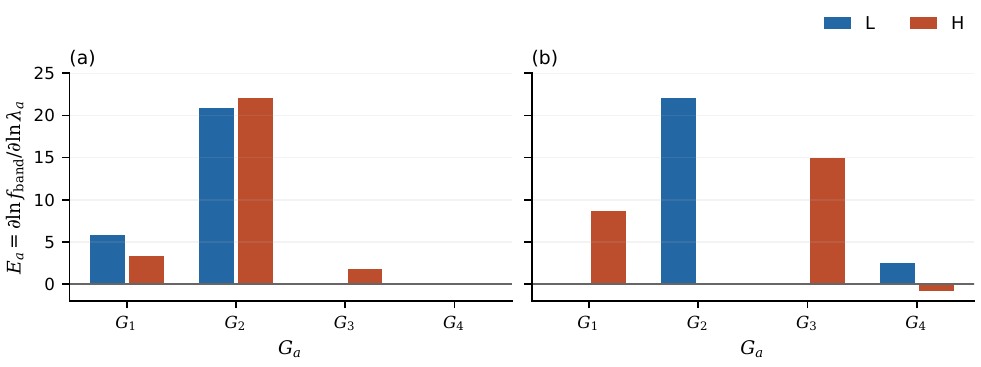}
\caption{Abundance response to the scalar groups for (a) the flexible QCD pair and (b) the stochastic curvature pair. The elasticities \(E_a\) are defined in Eq. \eqref{eq:A1}, with all other groups held fixed. For QCD H, the G2 elasticity is about 13 times larger than G3's, although their mutual coupling dominates the tensor signal at \(f_1\). The stochastic curvature derivatives include the covariance, spectral moments, peak radius and BBKS weight; a negative elasticity is a net response of these statistics, not a negative population.}
\label{fig:9}
\end{figure}

The group elasticities are compared in Figure~\ref{fig:9}. Each derivative rescales one group at fixed powers in the others, rather than following an abundance-compatible refit.

The Zenodo data \citep{yuzhoutang2026dataset} give the abundance residuals after removing G3, with and without restoring the total.

\FloatBarrier

\section{Formation definitions and numerical evaluation}\label{app:B}

Formation and tensor integrals are compared under domain and quadrature refinement and, where available, independent integration. Relative changes in the principal abundance bins and external responses are below \(3\times10^{-4}\), and the principal ideal-radiation window energies agree within \(3\times10^{-5}\). Refining the scalar-mode quadrature changes the completed QCD energies at the four comparison frequencies by less than 1.3\%. Comparisons at fixed primordial input are distinguished from those after solving again for the total abundance. The Zenodo archive \citep{yuzhoutang2026dataset} includes detailed numerical accuracy records for the formation and ideal-radiation calculations.

\subsection{QCD mass integration and absolute normalization}\label{sec:B.1}

Inverting the critical mass relation in Eq. \eqref{eq:6} gives the compaction excess \(\Delta_{\mathcal C}=\mathcal C-\mathcal C_{\rm th}\),

\stepcounter{equation}
\begin{equation}
\Delta_{\mathcal C}=\left(\frac{M}{\kappa M_H}\right)^{1/\gamma},\quad
D=1-\frac{\mathcal C_{\rm th}+\Delta_{\mathcal C}}{\Phi},\quad
g_M=2\Phi(1-\sqrt D).
\tag{B1}\label{eq:B1}
\end{equation}

For \(D>0\), \(dg_M/d\ln M=\Delta_{\mathcal C}/(\gamma\sqrt D)\). With \(\mu_H=M_H/M_\odot\), Eq. \eqref{eq:7} is equivalently

\stepcounter{equation}
\begin{equation}
\psi(M)=\int d\mu_H\,\mathcal T(\mu_H)
\frac{M}{M_H}\frac{\Delta_{\mathcal C}}{\gamma\sqrt D}
p(g_M;M_H)\,\Theta(D),
\tag{B2}\label{eq:B2}
\end{equation}

The Heaviside function \(\Theta(D)\) restricts the integrand to \(D>0\). Candidate QCD populations use \(p=p_{\rm pk}\). Threshold statistics, used only to construct the reference template, instead use

\stepcounter{equation}
\begin{equation}
p_G(g;\sigma_c^2)=\frac{1}{\sqrt{2\pi \sigma_c^2}}\exp\!\left(-\frac{g^2}{2\sigma_c^2}\right).
\tag{B3}\label{eq:B3}
\end{equation}

The conversion from formation fraction to present abundance is

\stepcounter{equation}
\begin{equation}
\begin{gathered}
\begin{aligned}
\mathcal T(\mu_H)&=\frac{\rho_E(T_H)a_H^3}{\Omega_{\rm DM}\rho_{c,0}c^2\mu_H}\\
&\simeq 3.76\times10^8\left(\frac{0.12}{\Omega_{\rm DM}h^2}\right)
\mu_H^{-3/2}\frac{[g_\rho(M_H)/106.75]^{3/4}}{g_s(M_H)/106.75}.
\end{aligned}
\end{gathered}
\tag{B4}\label{eq:B4}
\end{equation}

Here \(\rho_{c,0}=3H_0^2/(8\pi G)\), \(\Omega_{\rm DM}h^2=0.12\), \(h=0.674\), \(T_0=2.7255\) K and \(g_{s,0}=3.93\). Entropy conservation and dilution as \(a^{-3}\) give this conversion, with \(H^2=8\pi G\rho_E/(3c^2)\). The energy fraction \(\beta_H=\int(M_g/M_H)p\,dg\) includes collapsed mass efficiency. Converting the \(d\mu_H\) measure to \(d\ln M_H\) supplies \(\mu_H\) and recovers the usual \(M_H^{-1/2}\) scaling. The monochromatic limit agrees with \citep{carretal2021} under the same mass-efficiency convention.

The factor \(b_{\rm pk}=1\) in Eq. \eqref{eq:5} follows the adopted high-peak convention \citep{iovinoetal2024}. The alternative \(3^{-3/2}\) corresponds to the standard BBKS high-peak normalization with these spectral moments and horizon volume; it is not varied here. Neither is a unit-normalized probability weight.

The threshold comparison changes the Gaussian seed threshold, rather than multiplying \(\mathcal C_{\rm th}\) by the same factor,

\stepcounter{equation}
\begin{equation}
\begin{aligned}
g_{\rm th}&=2\Phi\left(1-\sqrt{1-\mathcal C_{\rm th}/\Phi}\right),\\
g_{\rm th}^{(\delta_{\rm th})}&=(1+\delta_{\rm th})g_{\rm th},\qquad
\mathcal C_{\rm th}^{(\delta_{\rm th})}=g_{\rm th}^{(\delta_{\rm th})}-\frac{[g_{\rm th}^{(\delta_{\rm th})}]^2}{4\Phi}.
\end{aligned}
\tag{B5}\label{eq:B5}
\end{equation}

We use \(\delta_{\rm th}=-0.15,-0.01,+0.01\), holding \(\kappa,\gamma,\Phi\) fixed and refitting to the original abundance bins, covariance, exact total and external curves. The constructed first-frequency ratios span about 15.2--15.6. This is a sensitivity comparison, rather than a probability distribution for formation uncertainty. Full spectra and responses are supplied with the data \citep{yuzhoutang2026dataset}.

\subsection{Conditional abundance template}\label{sec:B.2}

The reference curvature spectrum \(\mathcal P_0(k)\) has the lognormal form of Eq. \eqref{eq:13}. Its center \(k_0=5.7\times10^5\,\mathrm{Mpc}^{-1}\) is rounded from the candidate enhancement reconstructed in \citep{dingetal2026}. We adopt the width \(\Delta_0=0.60\) and integrated area \(\mathcal A_0=\sqrt{2\pi}P_{*,0}\Delta_0\simeq0.021\) for the smooth template. These parameters determine the generated mass-function shape before its abundance is assigned. The reconstruction for the LIGO--Virgo--KAGRA binary sample selected with a lower intrinsic black-hole mass cut of \(15\,M_\odot\) is tabulated at \(1,3,\ldots,101\,M_\odot\) \citep{dingwang2026data}. We use its two endpoints for our logarithmic mass bins. Threshold statistics in Eqs. \eqref{eq:B2} and \eqref{eq:B3} give \(\psi_{\rm th}[\mathcal P_0]\). We normalize this population by

\stepcounter{equation}
\begin{equation}
\psi_0(M)=\frac{f_{{\rm band},0}\,\psi_{\rm th}[\mathcal P_0](M)}
{\displaystyle\int_{M_-}^{M_+}\psi_{\rm th}[\mathcal P_0](M')\,d\ln M'}.
\tag{B6}\label{eq:B6}
\end{equation}

Integrating \(\psi_0\) gives every target bin and \(f_{{\rm all},0}\simeq7.25\times10^{-4}\). At this normalization its CMB response is about 0.81. For this fixed mass-function shape, Eq. \eqref{eq:12} makes the responses linear in the assigned band abundance, with the CMB curve allowing up to about \(1.2\times10^{-4}\). The chosen \(10^{-4}\) is one subdominant normalization within these adopted bounds. Changing \(\mathcal A_0\) also changes the shape of the generated mass function; the normalization in Eq. \eqref{eq:B6} does not fix that area. The normalization applies only to the synthetic target. QCD and stochastic-curvature candidates use their respective absolute formation conversions, Eqs. \eqref{eq:B4} and \eqref{eq:B11}.

The bin edges are \(M_j=M_-(M_+/M_-)^{(j-1)/N_M}\), \(j=1,\ldots,N_M+1\). At their geometric centers \(m_i=\sqrt{M_iM_{i+1}}\), the covariance is

\stepcounter{equation}
\begin{equation}
\begin{gathered}
\Sigma_{ij}=\varepsilon^2\left[
(1-\rho)\delta_{ij}+\rho
\exp\!\left(-\frac{\ln^2(m_i/m_j)}{2\ell^2}\right)\right].
\end{gathered}
\tag{B7}\label{eq:B7}
\end{equation}

Here \(\delta_{ij}\) is the Kronecker delta. We model the log-abundance errors as the sum of independent and smoothly correlated components. The choice \(\varepsilon=0.10\) gives a ten-percent marginal log error, and \(\rho=0.5\) assigns half of its variance to each component. We fix \(\ell\simeq0.37\) in \(\ln M\), using the abundance-weighted root-mean-square spread at fixed horizon mass from an earlier threshold reference calculation. This sets the scale of the assumed smooth errors; the covariance remains fixed for all candidates. The data give its precise value, construction and separate changes to the error model, together with the bin edges, targets and reference spectrum \citep{yuzhoutang2026dataset}.

\subsection{Single lognormal spectra}\label{sec:B.3}

For each lognormal shape, the exact total abundance fixes \(P_*\) through the QCD formation integral. The numerical search restricts \(\max_k\mathcal P_\zeta(k)\leq0.055\). Table~\ref{tab:3} gives the two fitted spectra.

\begin{table}[!htbp]
\centering
\small
\setlength{\tabcolsep}{4pt}
\renewcommand{\arraystretch}{1.18}
\begin{tabularx}{\textwidth}{@{}L>{\raggedleft\arraybackslash}X>{\raggedleft\arraybackslash}X@{}}
\toprule
Parameter or response & L & H \\
\midrule
Peak height \(P_*\) & 0.0132 & 0.0109 \\
Peak position \(k_*/\mathrm{Mpc}^{-1}\) & \(6.04\times10^5\) & \(5.28\times10^5\) \\
Logarithmic width \(\Delta\) & 0.424 & 0.816 \\
\(\chi^2_{\rm PBH}\) & 0.960 & 0.960 \\
\(\max_a\mathcal R_a\) & 0.708 & 0.917 \\
\(\omega(f_1)\) & \(1.9\times10^{-10}\) & \(5.0\times10^{-10}\) \\
\bottomrule
\end{tabularx}
\caption{Rounded parameters and responses for the single lognormal pair. Both spectra have the prescribed exact total. The maximum response is taken over all three external bounds, whose thresholds are one. The tensor energies use the common ideal-radiation normalization of Section 2.1. Unrounded parameters and predictions are supplied with the data \citep{yuzhoutang2026dataset}.}
\label{tab:3}
\end{table}

\begin{figure}[!htbp]
\centering
\includegraphics[width=\textwidth]{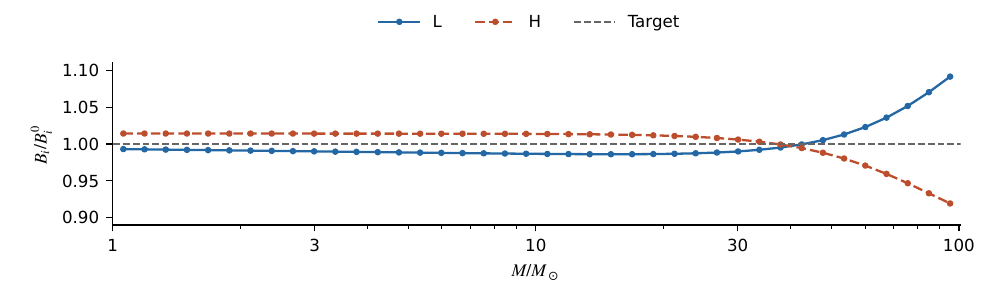}
\caption{Absolute single-lognormal abundance bins divided by the common target, with no normalization within the observed mass interval. The horizontal dashed line denotes the target; lines join the saved bin values. Both spectra satisfy the correlated abundance condition and the exact total. Colors identify L and H at \(f_1\).}
\label{fig:10}
\end{figure}

The source fractions in Section 4.1 insert \(\mathbf1_{q>k_*}\mathbf1_{p>k_*}\) into the complete tensor convolution, where \(q=ku\), \(p=kv\), and each spectrum uses its own peak position. To resolve formation, let \(dB_X(r_m)\) be the positive contribution to the observed-band abundance, with \(B_X=\int dB_X\). If \(\sigma_{c,>}^2\) restricts the variance integral to scalar modes above the peak, the weighted fraction is \(B_X^{-1}\int[\sigma_{c,>}^2(r_m)/\sigma_c^2(r_m)]\,dB_X(r_m)\). It gives 59\% and 71\% for L and H. This is a fraction of the variance at each formation radius, averaged over the abundance contribution, rather than a partition of PBH abundance by scalar wavenumber. All formation radii retain the complete spectrum. Replacing the scalar cut by other stated multiples of the peak gives the corresponding formation and tensor fractions. The data give these contributions for cuts from the peak to three times its wavenumber, as well as the separate formation-radius selections \citep{yuzhoutang2026dataset}.

\subsection{Flexible spectra and tensor quadrature}\label{sec:B.4}

The reference representation uses \(N_k=44\) components. One component has the reference feature's center and width; the other centers span \([10^5,10^7]\,\mathrm{Mpc}^{-1}\), with common width \(\Delta_{\rm b}=0.20\). Widths are standard deviations in \(\ln k\). This interval covers the reference scale and scalar scales contributing to the nanohertz tensor signal. The searches retain the power bound in Appendix~\ref{app:B}.3. Appendix~\ref{app:C}.2 varies the component count and width.

The single lognormal and the fixed dictionary are complementary spectral assumptions. The fixed centers and widths do not make this dictionary an exact superset of every lognormal in Eq. \eqref{eq:13}. Their constructed contrasts compare realizations of spectral freedom; no nesting of the two families is assumed.

In Eq. \eqref{eq:16}, component \(\alpha=0\) has \(x_0=\ln[k_0/(1\,\mathrm{Mpc}^{-1})]\) and width \(\Delta_0\). The other \(N_k-1\) centers are equally spaced in \(\ln k\) over the interval specified above, with width \(\Delta_{\rm b}\). Writing each unit-area component as \(G_\alpha\) gives \(\mathcal P_\zeta=\sum_\alpha c_\alpha G_\alpha\) and \(\omega(k)=\boldsymbol c^{\mathsf T}Q_k\boldsymbol c\). The entries of \(Q_k\) are the pair integrals. Analytic abundance and tensor derivatives are used to vary the component areas under the common conditions.

Table~\ref{tab:4} gives the abundance conditions and first-frequency energies of both flexible pairs.

\begin{table}[!htbp]
\centering
\small
\setlength{\tabcolsep}{4pt}
\renewcommand{\arraystretch}{1.18}
\begin{tabularx}{\textwidth}{@{}Lrrrr@{}}
\toprule
Condition or energy & QCD L & QCD H & Stochastic L & Stochastic H \\
\midrule
\(\chi^2_{\rm PBH}\) & 0.961 & 0.962 & 0.814 & 0.801 \\
Upper bound on \(\max\mathcal P_\zeta\) & 0.0177 & 0.0178 & 0.0235 & 0.0230 \\
\(\max_a\mathcal R_a\) & 0.990 & 0.990 & 0.990 & 0.818 \\
\(\omega(f_1)\) & \(4.6\times10^{-11}\) & \(7.1\times10^{-10}\) & \(1.9\times10^{-10}\) & \(2.1\times10^{-9}\) \\
\bottomrule
\end{tabularx}
\caption{Rounded constraint values and first-frequency energies for the flexible pairs. The maximum response is over all three external bounds. All populations have the prescribed exact total. Energies use the common ideal-radiation normalization of Section 2.1. Full individual responses and spectral comparisons are supplied with the data \citep{yuzhoutang2026dataset}.}
\label{tab:4}
\end{table}

The external curves are the baseline multi-frequency cosmic-radiation-background limits of Casanueva-Villarreal et al. \citep{casanuevavillarrealetal2025}, the two-standard-deviation Galactic-centre radio bound of Manshanden et al. \citep{manshandenetal2019}, and the conservative Park--Ricotti CMB feedback bound of Agius et al. \citep{agiusetal2024}. They are interpolated logarithmically within their tabulated mass supports, without extrapolation. Their adopted input tables and source versions are supplied with the data \citep{yuzhoutang2026dataset}.

The scalar moments include component tails beyond the center interval. The integration domains and full bin predictions are archived on Zenodo \citep{yuzhoutang2026dataset}.

The median horizon masses contributing to the observed band are about 240 and \(250\,M_\odot\) for L and H, whereas their CMB-weighted 90th percentiles are about 1200 and \(1700\,M_\odot\). Figure~\ref{fig:11} gives the corresponding cumulative weights for the stochastic curvature pair. Formation horizon mass differs from the final PBH mass on which an external bound is imposed.

The component-center estimate \(k_{\rm res}=(k_a+k_b)/\sqrt3\) lies inside the triangular tensor domain only for

\stepcounter{equation}
\begin{equation}
2-\sqrt3<\frac{k_a}{k_b}<2+\sqrt3.
\tag{B8}\label{eq:B8}
\end{equation}

This condition concerns the centers' resonance rather than an absolute cutoff for finite-width components \citep{caietal2019}.

The ideal radiation kernel in Eq. \eqref{eq:2}, with \(\Xi_{uv}=u^2+v^2-3\) and \(G=4v^2-(1+v^2-u^2)^2\), is

\stepcounter{equation}
\begin{equation}
\begin{aligned}
\mathcal K_{\rm RD}(u,v)&=\frac{3G^2\Xi_{uv}^2}{1024u^8v^8}
\Biggl\{
\left[-4uv+\Xi_{uv}\ln\left|\frac{3-(u+v)^2}{3-(u-v)^2}\right|\right]^2\\
&\qquad\qquad{}+\pi^2\Xi_{uv}^2\Theta(u+v-\sqrt3)
\Biggr\}.
\end{aligned}
\tag{B9}\label{eq:B9}
\end{equation}

The quadrature retains all cross terms and splits the logarithmic resonance. Continuous spectra and window integrals use direct tensor evaluation; spectral maxima and intersections use direct root finding. Their numerical accuracy is summarized in Appendix~\ref{app:B}.

\subsection{Stochastic curvature threshold and integration domain}\label{sec:B.5}

The second prescription follows the type I rare-peak construction of Germani and Sheth \citep{germanisheth2023}. It lets the candidate spectrum determine the joint Gaussian statistics of the primordial compaction seed \(g=-(4/3)r\partial_r\zeta\), its scale derivative \(v=r\partial_r g\) and its curvature \(w_{\rm pk}=-r^2\partial_r^2g\), where \(r\) is the comoving smoothing radius. Stochastic curvature refers to the distribution of peak shapes in a Gaussian field.

The Bardeen--Bond--Kaiser--Szalay (BBKS) spatial peak weight \citep{bardeenetal1986} is combined with the scale-maximum condition \(v=0,w_{\rm pk}>0\). A fixed radiation relation assigns a collapse threshold \(\mathcal C_c(w_{\rm pk})\) to each curvature \citep{escrivagermanisheth2020}. Applying the nonlinear radiation compaction relation \citep{youngmuscobyrnes2019} and critical mass scaling with a fixed coefficient converts the collapsing peaks into PBHs. Their mass-weighted comoving population gives \(\psi(M)\).

The Fourier kernel of the initial seed is \((4/3)krj_1(kr)\), where \(j_1\) is the spherical Bessel function. No horizon transfer is applied to these initial statistics. The moments are \(\sigma_j^2=\int d\ln k\,(kr)^{2j}[(4/3)krj_1(kr)]^2\mathcal P_\zeta(k)\). Differentiating the seed kernel gives the joint Gaussian density \(p(g,v,w_{\rm pk})\). The spatial and scale peak conditions give

\stepcounter{equation}
\begin{equation}
\frac{dn_{\rm pk}}{d\ln r\,dg\,dw_{\rm pk}}
=\frac{w_{\rm pk}\,f_{\rm BBKS}[(2g+w_{\rm pk})/\sigma_2]}{(2\pi R_*^2)^{3/2}}
p(g,v=0,w_{\rm pk}),\qquad R_*=\sqrt3\,r\frac{\sigma_1}{\sigma_2}.
\tag{B10}\label{eq:B10}
\end{equation}

Here \(n_{\rm pk}\) is comoving number density, \(R_*\) is the characteristic peak radius, and \(f_{\rm BBKS}\) is the full BBKS spatial peak function \citep{germanisheth2023,bardeenetal1986}. With \(\mathcal C(g)=g-3g^2/8\), the collapse range is \(g_c(w_{\rm pk})<g<4/3\), where \(g_c=(4/3)[1-\sqrt{1-3\mathcal C_c(w_{\rm pk})/2}]\). The final mass is \(M_g=6M_H[\mathcal C(g)-\mathcal C_c(w_{\rm pk})]^{0.36}\). Its mass-weighted population is

\stepcounter{equation}
\begin{equation}
\psi(M)=\frac{M}{\rho_{{\rm DM},0}}
\int d\ln r\int_0^\infty dw_{\rm pk}\int_{g_c(w_{\rm pk})}^{4/3}dg\,
\frac{dn_{\rm pk}}{d\ln r\,dg\,dw_{\rm pk}}\,
\delta_{\rm D}(\ln M-\ln M_g).
\tag{B11}\label{eq:B11}
\end{equation}

Because the number density is comoving with \(a_0=1\), this conversion needs no additional growth factor. The Friedmann and entropy relations define \(M_H(r)\) using fixed \(g_\rho=g_s=10.75\), the photon, electron--positron and coupled-neutrino content near the MeV era. Holding it fixed defines an ideal-radiation alternative, rather than an epoch-by-epoch thermal history.

Writing \(q\) for the dimensionless curvature parameter of the compaction profile, the radiation threshold is specified by \citep{escrivagermanisheth2020}

\stepcounter{equation}
\begin{equation}
\mathcal C_c(q)=\frac{4}{15}\frac{e^{-1/q}q^{1-5/(2q)}}{\gamma_{\rm inc}(5/(2q),1/q)},
\qquad w_{\rm pk}=4q\mathcal C_c(q)\sqrt{1-\tfrac32\mathcal C_c(q)}.
\tag{B12}\label{eq:B12}
\end{equation}

Here \(\gamma_{\rm inc}\) is the lower incomplete gamma function. Inverting the curvature relation gives \(\mathcal C_c(w_{\rm pk})\) and \(g_c(w_{\rm pk})=(4/3)[1-\sqrt{1-3\mathcal C_c(w_{\rm pk})/2}]\), in the radial convention of \citep{germanisheth2023}. The peak density is per comoving volume. Its absolute conversion uses \(\Omega_{\rm DM}h^2=0.12\), \(g_\rho=g_s=10.75\), \(T_0=2.7255\) K and \(g_{s,0}=3.93\).

\begin{figure}[!htbp]
\centering
\includegraphics[width=\textwidth]{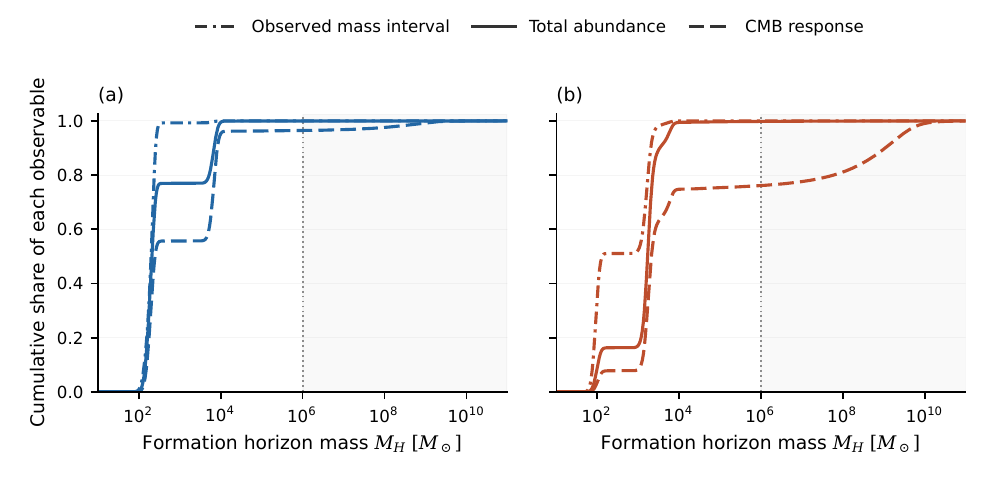}
\caption{Cumulative contributions to the observed-interval abundance, total abundance and CMB response for the stochastic curvature spectra: (a) L and (b) H. Each curve is divided by its own complete integral. The vertical line and shading mark \(M_H>10^6\,M_\odot\). The horizontal coordinate is formation horizon mass; abundance selection and external weights act on the final PBH mass after critical scaling.}
\label{fig:11}
\end{figure}

The horizon integral spans \(10^{-3}\)--\(10^{18}\,M_\odot\) in the formal radiation model. Figure~\ref{fig:11} compares the cumulative weights contributing to the observed-band abundance, total abundance and CMB response.

Matching each comoving radius to \(aH/c=1/r\) in a reference thermal background places the CMB contribution from \(M_H>10^6M_\odot\) mainly in the keV era. The formation integral retains the ideal-radiation \(M_H(r)\). Scales entering after the matter density reaches 1\% of the radiation density contribute less than \(4\times10^{-5}\) of either total, negligible observed-band abundance and an absolute CMB response below \(2\times10^{-7}\). The contribution after equality is smaller. The exact total includes the formal integration tail.

\subsection{Tensor strain, timing response and transfer}\label{sec:B.6}

In the tensor convention of \citep{kohriandterada2018}, the local radiative metric is \(ds^2=-c^2dt^2+[\delta_{ij}+h_{ij}/2]dx^i dx^j\). The physical spatial strain is \(h_{ij}/2\); its scalar-induced Newtonian representative is the gauge-invariant propagation strain of \citep{domenechpiwang2026}. The cosmological kernel in Eq. \eqref{eq:B9} already includes this metric normalization and the sum over tensor polarizations.

Let \(\tau_e,\tau_o\) be emitter and receiver proper times, \(u^\mu\) their four-velocities, and \(k_{\gamma,\mu}\) the photon wave covector. The fractional period shift is

\stepcounter{equation}
\begin{equation}
1+z=\frac{d\tau_o}{d\tau_e}
=\frac{(-u^\mu k_{\gamma,\mu})_e}{(-u^\mu k_{\gamma,\mu})_o}.
\tag{B13}\label{eq:B13}
\end{equation}

The contractions are scalars at the physical emission and reception events. The metric, light path and clock worldlines transform together when their coordinates change. We specialize the second-order timing expression of Magi and Yoo \citep{magiyoo2026timing}, derived on a Minkowski background, to a freely propagating tensor and geodesic clocks at rest in the background. For a pulsar at distance \(L\) in the unit direction \(\boldsymbol n\) from the observer, and a wave propagating along the unit vector \(\hat{\boldsymbol\Omega}\),

\stepcounter{equation}
\begin{equation}
\begin{aligned}
z_{\rm GW}(t)&=\frac{n^in^j}{4(1+\hat{\boldsymbol\Omega}\cdot\boldsymbol n)}
\left[h_{ij}(t,\boldsymbol0)-h_{ij}(t-L/c,L\boldsymbol n)\right],\\
r_{\rm GW}(t)&=\int^t z_{\rm GW}(t')\,dt'.
\end{aligned}
\tag{B14}\label{eq:B14}
\end{equation}

The pulsar term samples the wave at phase \(t-(L/c)(1+\hat{\boldsymbol\Omega}\cdot\boldsymbol n)\). The response is linear in the generated strain, which is already second order in the primordial scalar field. Only its time variation contributes to the propagation redshift \citep{domenechpiwang2026}.

For an isotropic, unpolarized, stationary tensor background, the single-sided auto-spectra in the long-distance limit, including incoherent Earth and pulsar terms, are

\stepcounter{equation}
\begin{equation}
S_r(f)=\frac{H_0^2\Omega_{{\rm GW},0}(f)}{8\pi^4f^5}
=\frac{(H_0/h)^2\omega(f)}{8\pi^4f^5},
\qquad S_z(f)=(2\pi f)^2S_r(f).
\tag{B15}\label{eq:B15}
\end{equation}

A band contributes \(\int_{\mathcal B}S_r(f)\,df\) to the residual variance. Its energy spectrum is therefore weighted by \(f^{-4}\) in the measure \(d\ln f\). For the principal pair, the two window variance ratios are about 0.95 and 44 before timing-model fitting. The higher-window residual RMS values are about 9 ns and 60 ns under the common ideal-radiation energy conversion of Section 2.1.

For the finite-time comparison, let \(\Pi_2\) be the orthogonal projection onto \(1,t,t^2\) on a centered interval of duration \(T=16.03\) Julian years with continuous uniform sampling. Define \(g_f(t)=(1-\Pi_2)e^{2\pi ift}\). The projected signal covariance is \citep{allenetal2026legendre}

\stepcounter{equation}
\begin{equation}
C^{\rm post}_r(t,t')=\int_0^\infty df\,S_r(f)
\operatorname{Re}[g_f(t)g_f^*(t')].
\tag{B16}\label{eq:B16}
\end{equation}

The time-averaged diagonal is the integral of \(S_r(f)[1-\sum_{\ell=0}^2(2\ell+1)j_\ell^2(\pi fT)]\) over input frequencies, with \(j_\ell\) the spherical Bessel functions. The broadband RMS is the square root of this integral over 0.01 to 300 nHz. The principal pair gives about 360 ns and 420 ns, a ratio of 1.17. Restricting the same integral to \(\mathcal B_2\) gives a variance ratio of about 45. Dividing by the broadband variance ratio before taking the square root gives 5.8 for the higher-window RMS ratio at equal broadband RMS. The window labels the input frequencies contributing to the nonstationary fitted covariance.

The transfer calculation uses conformal length \(\eta=\int c\,dt/a\) in Mpc and \(\mathcal H=a'/a=aH/c\); primes denote \(\eta\) derivatives. Let \(w_{\rm bg}\) be the background equation of state, \(c_s^2\) the squared sound speed in units of \(c^2\), and \(T_\Psi(q,\eta)\) the potential transfer for scalar wavenumber \(q\). They obey \citep{abeetal2021}

\stepcounter{equation}
\begin{equation}
\begin{aligned}
w_{\rm bg}&=\frac{4g_s}{3g_\rho}-1,\qquad
c_s^2=\frac{4g_s(4+d\ln g_s/d\ln T)}{3g_\rho(4+d\ln g_\rho/d\ln T)}-1,\\
T_\Psi''(q,\eta)&+3\mathcal H(1+c_s^2)T_\Psi'(q,\eta)
+[c_s^2q^2+3\mathcal H^2(c_s^2-w_{\rm bg})]T_\Psi(q,\eta)=0.
\end{aligned}
\tag{B17}\label{eq:B17}
\end{equation}

The potential transfer tends to one in early radiation, when the primordial potential is \(-2\zeta/3\). For scalar wavenumbers \(q,p\), the pair source enters the transverse-traceless source \(S_k\) of tensor mode \(h_k\). The pair source factor and tensor evolution equation are

\stepcounter{equation}
\begin{equation}
\begin{gathered}
2T_\Psi(q,\eta)T_\Psi(p,\eta)+\frac{4}{3(1+w_{\rm bg})}
\left(T_\Psi(q,\eta)+\frac{T_\Psi'(q,\eta)}{\mathcal H}\right)
\left(T_\Psi(p,\eta)+\frac{T_\Psi'(p,\eta)}{\mathcal H}\right),\\
(a h_k)''+\left[k^2-\frac{1-3w_{\rm bg}}{2}\mathcal H^2\right](a h_k)=4aS_k.
\end{gathered}
\tag{B18}\label{eq:B18}
\end{equation}

The background follows continuity and Friedmann evolution with entropy normalization at 5 MeV; below that temperature it joins pure radiation. This matching point is after the QCD crossover and before the MeV-era neutrino and electron--positron evolution omitted from this solver. We convert the asymptotic tensor fraction of this matched radiation model using the thermal degrees of freedom at the reference temperature \(T_m=5\,\mathrm{MeV}\), with present-day conversion \citep{cecchinietal2025,abeetal2021} \(\mathcal T_{\rm GW}(T_m)=\Omega_{r,0}h^2[g_\rho(T_m)/g_{\rho,0}][g_{s,0}/g_s(T_m)]^{4/3}\). The tabulated degrees of freedom at the matching temperature give \(\mathcal T_{\rm GW}(5\,\mathrm{MeV})=3.48\times10^{-5}\), which is 2.13 times the principal reference factor. The outgoing tensor is extracted from the two source projections onto a common sine--cosine homogeneous basis, including the remaining source after the numerical evolution. In the adopted radiation background below the matching point, set \(x=k\eta\), \(\tilde x=x-x_m+k/\mathcal H_m\), where \(x_m\) and \(\mathcal H_m\) refer to that point. The potential transfer has the exact continuation

\stepcounter{equation}
\begin{equation}
T_\Psi(q,\tilde x)=A_q\frac{\sin y-y\cos y}{y^3}
+B_q\frac{\cos y+y\sin y}{y^3},
\qquad y=\frac{q\tilde x}{\sqrt3\,k}.
\tag{B19}\label{eq:B19}
\end{equation}

The coefficients \(A_q,B_q\) are fixed by the transfer and its derivative at the numerical handoff. Products in Eq. \eqref{eq:B18} reduce the remaining source projections to oscillatory inverse powers, which are integrated to infinite time using exponential integrals, including the acoustic-resonance terms. The oscillation-averaged outgoing power uses the squared norm of the completed two-component projection. Continuous spectra and the four comparison frequencies use this same late-time prescription.

\begin{figure}[!htbp]
\centering
\includegraphics[width=\textwidth]{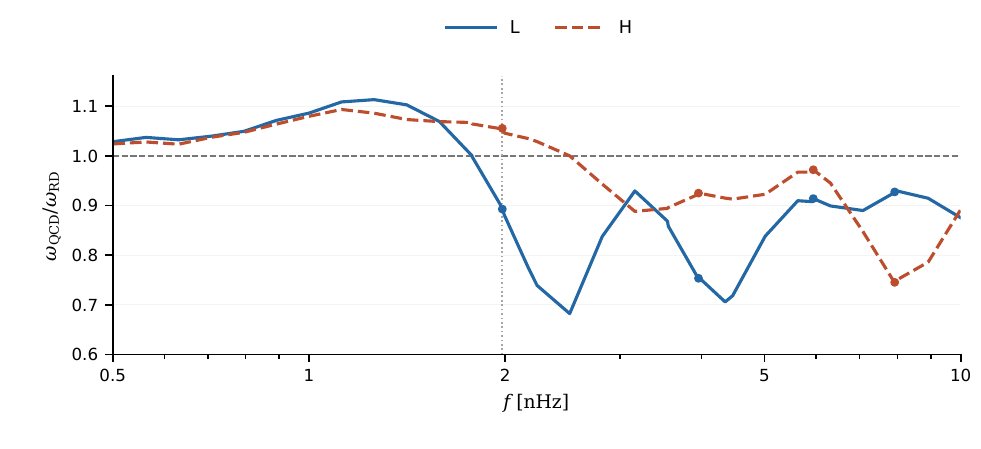}
\caption{QCD-to-ideal-radiation tensor-energy ratios for the unchanged flexible QCD pair at a common redshift normalization. The horizontal line denotes equal energy, the vertical line marks \(f_1\), and dots mark the four directly evaluated comparison frequencies. The ratios use the completed late-time source integrals and isolate scalar, source and tensor evolution: the factor of 2.13 between the native redshift conversions has been divided out. The primordial spectra and PBH predictions are unchanged.}
\label{fig:12}
\end{figure}

Holding the flexible QCD spectra and their PBH predictions fixed, QCD scalar, source and tensor evolution raises the first-frequency ratio from about 15.5 to 18 (Figure~\ref{fig:12}). On the common redshift reference, it lowers L by about 11\% and raises H by about 6\%. Using the alternative thermal input of Saikawa and Shirai \citep{saikawaandshirai2018} changes the first-frequency ratio by less than 1\%.

For the separate neutrino diffusion prescription, the ideal-radiation transfer \(T_\Psi^{\rm RD}\) is damped as \(T_\Psi^{\rm diff}=T_\Psi^{\rm RD}\exp[-D_q]\), following the shear-viscosity approximation of \citep{yuetal2025,domenechandchluba2025}. Let \(f_\nu\) be the neutrino fraction of radiation energy, and let the subscript dec denote neutrino decoupling. In a tightly coupled radiation fluid, \(d k_D^{-2}/d\eta\simeq2\eta_\nu/[3a(\rho+P)]\), with \(\eta_\nu=(4/15)\rho_\nu t_\nu\). We approximate weak scattering by \(t_\nu\propto T^{-5}\propto a^5\), set \(t_{\nu,\rm dec}=H_{\rm dec}^{-1}\), and use \(a\propto\eta\) and \(a_{\rm dec}H_{\rm dec}/c=\eta_{\rm dec}^{-1}\). The integral then gives \(k_D^{-2}(\eta)=(2f_\nu/75)k_{\rm hor,dec}^{-2}(\eta/\eta_{\rm dec})^5\) before decoupling. We stop the fluid diffusion at decoupling, giving the implemented piecewise prescription

\stepcounter{equation}
\begin{equation}
D_q(\eta)=\left(\frac{q}{k_{D,\rm dec}}\right)^2
\min\!\left[\left(\frac{\eta}{\eta_{\rm dec}}\right)^5,1\right],
\qquad
k_{D,\rm dec}=\frac{k_{\rm hor,dec}}{\sqrt{2f_\nu/75}}.
\tag{B20}\label{eq:B20}
\end{equation}

We use \(T_{\rm dec}=1.5\) MeV as a representative instantaneous neutrino-decoupling point \citep{yuetal2025}, not a measured sharp transition. The ideal-fluid degrees of freedom are \(g_*=10.75=2+3.5+5.25\) for photons, electron--positron pairs and three neutrino species, so \(f_\nu=5.25/10.75\). The Friedmann and entropy relations set \(k_{\rm hor,dec}\) and Eq. \eqref{eq:B20} sets \(k_{D,\rm dec}\). The coefficient in Eq. \eqref{eq:B20} follows from the stated approximations, including the chosen decoupling condition; it is not a fully calibrated neutrino kinetic transfer. The derivative \((T_\Psi^{\rm diff})'=[(T_\Psi^{\rm RD})'-D_q'T_\Psi^{\rm RD}]e^{-D_q}\) enters the velocity source with the transfer itself. The integration settings and point and window energies are supplied with the data \citep{yuzhoutang2026dataset}.

The separate scalar-diffusion prescription gives a first-frequency ratio near 12 for the unchanged flexible QCD pair. It changes the lognormal pair's second-window ratio from about 58 to 53. For the stochastic-curvature pair, the first-frequency ratio falls from 11 to 8.4 and the second-window ratio changes from 0.055 to 0.064, retaining the reversed ordering. These comparisons retain the leading-order tensor source; the data provide the point and window energies \citep{yuzhoutang2026dataset}.

\subsection{Window responses along the lognormal branch}\label{sec:B.7}

The lognormal shape coordinates are \(\boldsymbol\theta=(\ln[k_*/(1\,\mathrm{Mpc}^{-1})],\ln\Delta)\). At every shape, the amplitude is solved from the exact total before differentiating. Partial derivatives hold the other independent coordinates fixed. The resulting amplitude and bin responses are

\stepcounter{equation}
\begin{equation}
\begin{aligned}
\frac{d\ln A}{d\theta_j}&=-\frac{\partial_{\theta_j}\ln f_{\rm all}}{\partial_{\ln A}\ln f_{\rm all}},\\
\widetilde J_{ij}\equiv\frac{d\ln B_i}{d\theta_j}
&=\partial_{\theta_j}\ln B_i+\partial_{\ln A}\ln B_i\,\frac{d\ln A}{d\theta_j}.
\end{aligned}
\tag{B21}\label{eq:B21}
\end{equation}

For a fixed window \(\mathcal B_j\), write \(\overline\omega_j=A^2\overline Q_j\). The derivative of its logarithmic average is energy weighted,

\stepcounter{equation}
\begin{equation}
\begin{aligned}
\overline Q_j&=\frac{\int_{\mathcal B_j}Q_f\,d\ln f}{\int_{\mathcal B_j}d\ln f},\\
\nabla\ln\overline Q_j
&=\frac{\int_{\mathcal B_j}Q_f\nabla\ln Q_f\,d\ln f}
{\int_{\mathcal B_j}Q_f\,d\ln f}.
\end{aligned}
\tag{B22}\label{eq:B22}
\end{equation}

Let \(\boldsymbol\theta_0\) be the reference abundance fit, \(\delta\boldsymbol\theta=\boldsymbol\theta-\boldsymbol\theta_0\), and \(L\) the lower triangular Cholesky factor of the covariance, \(LL^{\mathsf T}=\Sigma\). Define \(\boldsymbol e=L^{-1}(\ln\boldsymbol B-\ln\boldsymbol B^0)\), \(\boldsymbol e_0=\boldsymbol e(\boldsymbol\theta_0)\), \(J_{\rm w}=\left.\partial\boldsymbol e/\partial\boldsymbol\theta\right|_0=L^{-1}\widetilde J(\boldsymbol\theta_0)\), and \(\boldsymbol d_j=\left.\nabla_{\boldsymbol\theta}\ln\overline\omega_j\right|_0\). For full column rank \(J_{\rm w}\), define \(\Lambda_j^2=\boldsymbol d_j^{\mathsf T}(J_{\rm w}^{\mathsf T}J_{\rm w})^{-1}\boldsymbol d_j\), and let \(J_{\rm w}^{+}\) denote the Moore--Penrose pseudoinverse. The component \(\boldsymbol e_\perp=(I-J_{\rm w}J_{\rm w}^{+})\boldsymbol e_0\) cannot be changed by a linearized shape variation. Completing the square bounds the first-order logarithmic window change \(\boldsymbol d_j^{\mathsf T}\delta\boldsymbol\theta\) as

\stepcounter{equation}
\begin{equation}
\begin{aligned}
\|\boldsymbol e_0+J_{\rm w}\delta\boldsymbol\theta\|^2&\leq1,\\
\left|\boldsymbol d_j^{\mathsf T}\delta\boldsymbol\theta+\boldsymbol d_j^{\mathsf T}J_{\rm w}^{+}\boldsymbol e_0\right|
&\leq\Lambda_j\sqrt{1-\|\boldsymbol e_\perp\|^2}.
\end{aligned}
\tag{B23}\label{eq:B23}
\end{equation}

The linearized region is empty if \(\|\boldsymbol e_\perp\|>1\). Write \(J_{\rm w}=U\operatorname{diag}(s_\alpha)V^{\mathsf T}\), with right singular vectors \(\boldsymbol v_\alpha\). Amplitude and pair responses contribute through their signed sum,

\stepcounter{equation}
\begin{equation}
\begin{aligned}
\Lambda_j^2&=\sum_\alpha\left[
\frac{2\boldsymbol v_\alpha\cdot\nabla\ln A}{s_\alpha}
+\frac{\boldsymbol v_\alpha\cdot\nabla\ln\overline Q_j}{s_\alpha}\right]^2,\\
\nabla\ln\frac{\overline\omega_2}{\overline\omega_1}
&=\nabla\ln\frac{\overline Q_2}{\overline Q_1}.
\end{aligned}
\tag{B24}\label{eq:B24}
\end{equation}

All gradients in Eq. \eqref{eq:B24} are evaluated at the reference fit. The common amplitude cancels exactly in the ratio response. The abundance Jacobian has rank two, with one shape direction much less constrained than the other. A rank-deficient abundance map gives an unbounded response under the linearized abundance condition if the tensor gradient has a component in its null space.

At the reference fit, the local responses are \(\Lambda_1\simeq0.093\) and \(\Lambda_2\simeq1.6\). The amplitude term partly offsets the lower-window pair response, while the higher window retains most of it. The two singular values differ by a factor of about 69. The finite abundance-compatible branch is not described by one reference tangent; its sampled derivatives and linearization residuals are supplied with the data \citep{yuzhoutang2026dataset}. The finite comparison of L and H is resolved exactly as

\stepcounter{equation}
\begin{equation}
\ln\frac{\overline\omega_{H,j}}{\overline\omega_{L,j}}
=2\ln\frac{A_H}{A_L}+\ln\frac{\overline Q_{H,j}}{\overline Q_{L,j}}.
\tag{B25}\label{eq:B25}
\end{equation}

This is the finite counterpart of the compensation in Eq. \eqref{eq:9}.

For a controlled formation comparison, we also use joint threshold statistics \citep{iovinoetal2024} with transferred and superhorizon Gaussian moments \citep{young2022ng}. The common target, covariance, total, collapse tables, mass mapping and external curves are unchanged. The variance of the linear compaction agrees with the high-peak calculation to numerical precision; the first comparison changes the counting measure. The second holds the threshold measure fixed while removing the radiation transfer from the Gaussian moments. The lower-window energies of the three Gaussian reference fits span a factor of about 2.2; their parameters and local responses are supplied with the data \citep{yuzhoutang2026dataset}.

The fixed-input comparison evaluates each reference spectrum under all three operators without readjusting its amplitude. The changes are not common rescalings of abundance. Refitting restores the common conditions but changes the conditional tensor prediction. The tested formation treatments retain the window asymmetry; their lower-window normalization varies more than the roughly ten-percent shifts induced by \(f_{\rm NL}=\pm0.1\) within the superhorizon threshold treatment below. This comparison is a spread among stated prescriptions, not a calibrated formation uncertainty. For the Gaussian lognormal pair on the superhorizon threshold branch, the same abundance conditions permit lower- and higher-window ratios of about 1.01 and 4.9.

\subsection{A local non-Gaussian comparison}\label{sec:B.8}

The local quadratic model is \(\zeta=G+(3/5)f_{\rm NL}(G^2-\langle G^2\rangle)\) \citep{byrnescopelandgreen2012}, with a lognormal spectrum for the Gaussian seed \(G\). Its peak height is \(A\), with the shape in Eq. \eqref{eq:13}. Gaussian and non-Gaussian candidates are both evaluated with the same joint threshold treatment and superhorizon moments. For the jointly Gaussian linear compaction \(X\) and shell curvature \(Z\), the spherical local prescription \citep{iovinoetal2024,young2022ng} gives

\stepcounter{equation}
\begin{equation}
\begin{aligned}
\mathcal C_\ell&=X\left(1+\frac65 f_{\rm NL}Z\right),\qquad
\mathcal C=\mathcal C_\ell-\frac{\mathcal C_\ell^2}{4\Phi},\\
p_\ell(l)&=\int dX\,dZ\,p_G(X,Z)\,
\delta_{\rm D}\!\left[l-X\left(1+\frac65f_{\rm NL}Z\right)\right].
\end{aligned}
\tag{B26}\label{eq:B26}
\end{equation}

Here \(p_G\) is the joint Gaussian density of \(X,Z\). We integrate both transformation roots with the absolute Jacobian, then use the shared type I collapse domain and critical mass mapping. This common spherical profile approximation is distinct from the high-peak measure in Eq. \eqref{eq:5}; it does not supply a new collapse calibration for non-Gaussian profiles. The search restriction in Appendix~\ref{app:B}.3 applies to the complete curvature power, including the quadratic convolution.

For the second-order tensor source, we retain all disconnected and connected four-point contributions \citep{unal2019,adsheadetal2021ng}. Gaussian parity makes the dependence on the local parameter even at fixed seed amplitude and shape,

\stepcounter{equation}
\begin{equation}
\begin{aligned}
\omega(f)&=A^2Q_{0f}+\left(\frac35 f_{\rm NL}\right)^2A^3Q_{2f}
+\left(\frac35 f_{\rm NL}\right)^4A^4Q_{4f},\\
\left.\frac{d\ln\overline\omega_j}{df_{\rm NL}}\right|_{\boldsymbol\theta,f_{\rm all},\,f_{\rm NL}=0}
&=2\left.\frac{d\ln A}{df_{\rm NL}}\right|_{\boldsymbol\theta,f_{\rm all},\,f_{\rm NL}=0}.
\end{aligned}
\tag{B27}\label{eq:B27}
\end{equation}

The coefficients \(Q_{nf}\) depend on the seed shape and the radiation kernel, with \(Q_{0f}=Q_f\). At \(f_{\rm NL}=0\), the two window energies have the same first-order fractional response when the seed shape and total abundance are held fixed. Weak non-Gaussianity at fixed seed shape moves the two window energies almost together after matching the total. At \(f_{\rm NL}=\pm0.1\), an amplitude adjustment alone fails the bin conditions; refitting peak position and width restores feasibility and gives a larger relative change in the higher window. Separate pairs at fixed \(f_{\rm NL}=-0.2,-0.1,+0.1,+0.2\) have lower-window contrasts of at most about one percent and higher-window contrasts of about 2--6.5. Individual cases are supplied with the data \citep{yuzhoutang2026dataset}.

For the four representative refits at \(f_{\rm NL}=\pm0.1,\pm0.2\), the explicit \(f_{\rm NL}^2\) and \(f_{\rm NL}^4\) terms in Eq. \eqref{eq:B27} contribute less than 0.4\% of either window energy. The disconnected and connected terms are reported separately in the data. A Gaussian field assigned the same complete curvature power, including the quadratic convolution, has a tensor spectrum equal to the disconnected contribution. Keeping that power fixed, its total PBH abundance is about 2.2 and 0.17 times the prescribed total for the refitted \(f_{\rm NL}=-0.1\) and \(+0.2\) examples, respectively.

Four feasible Gaussian lognormals reproduce the ratios of the tested non-Gaussian refits' window energies without a free tensor normalization. Their absolute window energies and full-band shapes remain different. The window ratio is therefore degenerate between these statistical and spectral choices. The examples are separate refits at fixed \(f_{\rm NL}\).

\FloatBarrier

\section{Abundance information and spectral assumptions}\label{app:C}

\subsection{Exterior mass information and CMB bounds}\label{sec:C.1}

The mass coverage study in Section 4.5 adds bins up to the abundance-weighted median and then the 90th percentile while retaining the original bins. The quantiles integrate \(\psi(M)\,d\ln M/f_{{\rm all},0}\) in final PBH mass. The target bins and covariance matrices are archived on Zenodo \citep{yuzhoutang2026dataset}.

The covariance retains Eq. \eqref{eq:B7}, with the same errors and correlation kernel evaluated at all bin centers. Its original principal block is unchanged. For original and added residuals \(\boldsymbol r_o,\boldsymbol r_a\), the block inverse decomposes the extended abundance distance into the original distance plus the nonnegative conditional quadratic form with covariance \(\Sigma_{aa}-\Sigma_{ao}\Sigma_{oo}^{-1}\Sigma_{oa}\). The joint tolerance therefore defines nested feasible sets as mass information is added. The total abundance, external curves and numerical search domain remain fixed within this experiment. The pairs in Table~\ref{tab:2} are selected at \(f_1\); the data also provide separate constructions selected by each window energy.

At fixed \(M_H\), the critical mass relation in Eq. \eqref{eq:6} gives \(dg/d\ln M\propto M^{1/\gamma}\) near threshold, while the final-mass weight contributes \(M\). Treating the remaining peak density as constant in the threshold excess gives \(\psi(M\mid M_H)\propto M^{1+1/\gamma}\). Rare-peak exponential factors and the varying QCD calibration limit that approximation at finite mass. For the directly generated reference population, the leading asymptote overestimates the original band abundance by about 37\% and changes the normalized bin shape by as much as 27\%. The original interval therefore already carries shape information.

For the flexible class, a separate extension adds three high-mass bins and then two low-mass bins. Their targets come from the initial feasible spectrum closest to the original abundance bins and are fixed before tensor optimization. The added bins have independent log errors of 0.10. The original targets and covariance block, total abundance and external bounds are retained under the same joint tolerance. The first-frequency contrast decreases from 15.5 to about 10.6 with the high-mass bins and to 10.5 after adding the low-mass bins. The data provide the added bin edges, targets, spectra and constraint values \citep{yuzhoutang2026dataset}.

To change the CMB curve's mass dependence while preserving its response to the reference population, define

\stepcounter{equation}
\begin{equation}
\begin{gathered}
f_{\max,\alpha_{\rm CMB}}(M)=A_{\alpha_{\rm CMB}} f_{\max,0}(M)
\left(\frac{M}{100\,M_\odot}\right)^{\alpha_{\rm CMB}},\\
\alpha_{\rm CMB}=\pm0.3,\quad 10\leq M/M_\odot\leq10^4,
\end{gathered}
\tag{C1}\label{eq:C1}
\end{equation}

\stepcounter{equation}
\begin{equation}
A_{\alpha_{\rm CMB}}=
\frac{\displaystyle\int_{10M_\odot}^{10^4M_\odot}d\ln M\,
\frac{\psi_0(M)}{f_{\max,0}(M)}
\left(\frac{M}{100M_\odot}\right)^{-\alpha_{\rm CMB}}}
{\displaystyle\int_{10M_\odot}^{10^4M_\odot}d\ln M\,
\frac{\psi_0(M)}{f_{\max,0}(M)}}.
\tag{C2}\label{eq:C2}
\end{equation}

Here \(f_{\max,0}\) is the adopted CMB curve. By construction, both tilted curves give the same CMB response for \(\psi_0\) as the adopted curve. Refitting the flexible spectra under the two tilted curves gives first-frequency contrasts of about 15 and 17. Removing the CMB condition gives a contrast near 39, but those two diagnostic populations violate the original curve by about 98 and 138 times. Their responses under both curves are supplied with the data \citep{yuzhoutang2026dataset}.

Uniform tightening instead replaces \(f_{\max,\rm CMB}\) by \(f_{\max,\rm CMB}/t\). The reference response is about 0.81, so tightening by about 1.24 excludes that complete reference population. It does not exclude all populations matching its observed-band abundances and total: an abstract nonnegative mass function can satisfy them with a CMB response about 0.060 at \(t=2\). Its exterior abundance lies at smaller final masses than in the exhibited formation spectra. The recorded formation searches do not establish whether the specified spectral family contains such a solution. The reference response, abstract population and search results are supplied with the data \citep{yuzhoutang2026dataset}.

\subsection{Coefficient count and component width}\label{sec:C.2}

Representation comparisons retain the common abundance target, covariance, exact total and external curves. The reference component and the wavenumber interval of the other centers remain as specified in Appendix~\ref{app:B}.4. Reducing the component count from 44 to 22 retains a first-frequency contrast of about 13; increasing it to 66 gives about 17. These representations are not assumed to be nested. Section 4.3 gives the width dependence. All component coefficients, energies and constraint values are supplied with the data \citep{yuzhoutang2026dataset}.

The resonance neighborhood is \(\left|\ln[(k_a+k_b)/(\sqrt3k_1)]\right|<\delta\), with actual integration wavenumbers \(k_a=k_1u\), \(k_b=k_1v\). For \(\delta=0.05,0.10,0.20\), it contains about 75\%, 91\% and 99\% of L's energy increase from width 0.20 to 0.40. The data give the remaining group and formation-response diagnostics \citep{yuzhoutang2026dataset}.

\subsection{Broken-power-law spectra}\label{sec:C.3}

A broken power law provides a separate test of single-peak morphology with independent edges. We use the convention \citep{iovinoetal2024}

\stepcounter{equation}
\begin{equation}
\mathcal P_\zeta(k)=A\frac{\alpha+\beta}
{\beta(k/k_*)^{-\alpha}+\alpha(k/k_*)^\beta},\qquad
\alpha,\beta>0.
\tag{C3}\label{eq:C3}
\end{equation}

The smoothing parameter is fixed to unity, \(\alpha\) and \(\beta\) are the rising and declining asymptotic slopes, \(A\) is the peak height. With both slopes restricted to \([0.2,4]\), direct QCD formation evaluation under the same target, covariance, exact total and external curves, with the power bound in Appendix~\ref{app:B}.3, gives two spectra satisfying the common conditions with a first-frequency ratio of 2.9. This comparison retains a comparable contrast without lognormal symmetry; the energies and parameters of both spectra are archived \citep{yuzhoutang2026dataset}.

The low candidate reaches the allowed declining slope \(\beta=4\), and the high candidate reaches the rising slope \(\alpha=4\). The edge domain is therefore part of the result. Broader slope ranges are reported in the supporting data but are not used for the principal claims here. This is a QCD-formation comparison. The untransferred \(\sigma_2^2\) moment of the stochastic curvature rule would require a declining asymptotic slope \(\beta>4\), or an explicit ultraviolet completion, so these BPL candidates are not inserted into that rule.

\subsection{Dependence on the prescribed target and covariance}\label{sec:C.4}

Covariance variations retain the target bins, exact total, external curves and frequencies. Target variations retain \(f_{{\rm band},0}=10^{-4}\) and the generating feature's integrated area; normalizing the mass function through Eq. \eqref{eq:B6} changes the all-mass total. Other inputs retain their principal values. Table~\ref{tab:5} shows representative window contrasts; complete cases and absolute energies are supplied with the data \citep{yuzhoutang2026dataset}.

\begin{table}[!htbp]
\centering
\small
\setlength{\tabcolsep}{4pt}
\renewcommand{\arraystretch}{1.18}
\begin{tabularx}{\textwidth}{@{}Lrr@{}}
\toprule
Abundance information & \(\overline F_1\) & \(\overline F_2\) \\
\midrule
Reference refit & 1.2 & 62 \\
Uncorrelated errors, \(\rho=0\) & 1.1 & 11 \\
Smaller marginal error, \(\varepsilon=0.05\) & 1.1 & 5.2 \\
Reference center \(1.15k_0\) & 1.1 & 6.2 \\
Reference width \(\Delta_0=0.70\) & 1.1 & 8.5 \\
\bottomrule
\end{tabularx}
\caption{Ratios of separately averaged window energies for representative changes to the abundance information. The reference row is a separate fit under the principal information model. The spectra are selected at \(f_1\); each window ratio is a constructive lower bound for that information model. Energies use the common ideal-radiation normalization and the windows in Eq. \eqref{eq:15}.}
\label{tab:5}
\end{table}

The higher-window contrast changes much more than the lower-window contrast. Removing correlations or tightening the marginal error compresses the displayed difference, and changing the target also changes its magnitude. The frequency redistribution persists for these constructions, with its size set by the abundance information.

\FloatBarrier
\begingroup
\small\fontsize{10}{10.8}\selectfont

\endgroup

\end{document}